# Application of Carbon Nanomaterials in Electronics Industry

Joydip Sengupta

Department of Electronic Science, Jogesh Chandra Chaudhuri College, Kolkata-700033

E-mail: joydipdhruba@gmail.com

## 1. Introduction

Nanomaterials have much improved property compared to their bulk counterparts which promote them as ideal material for applications in different industry. Among various nanomaterials, the different nano-allotropes of carbon viz. fullerene, carbon nanotube and graphene (Fig.1) are the most important ones as being the fact that all of their discoverers are noted with prestigious awards like Nobel prize or Kavil prize. Carbon forms different nano-allotropes by varying the nature of orbital hybridisation. Since all the nano-allotropes of carbon possess exotics physical and chemical properties thus they are extensively used in different application especially in electronics industry.

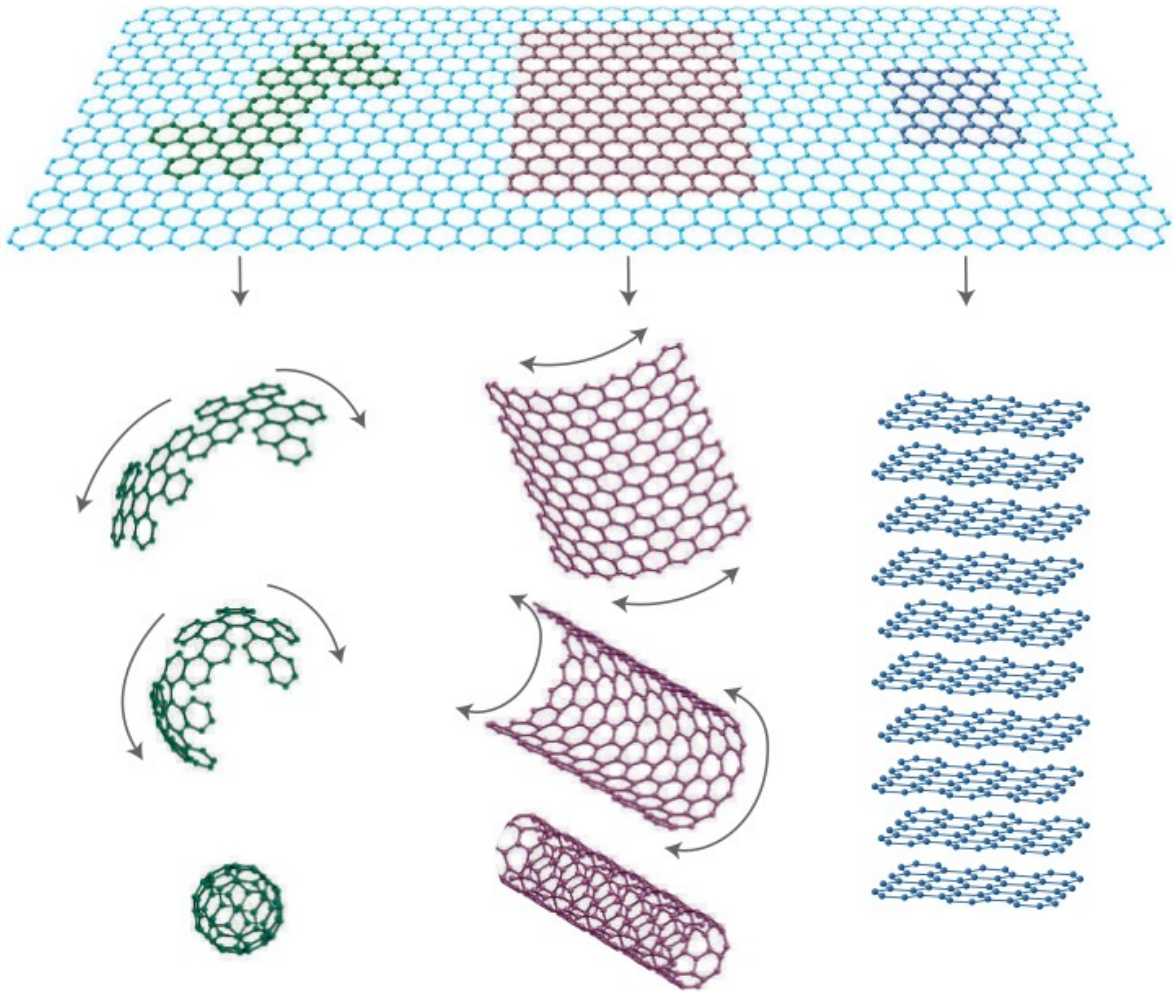

*Fig.1. Graphene is a 2D material that acts as a building block for carbonaceous materials of all other dimensions[1].(Reprinted with permission from Springer Nature)*

## 2. Fullerene

The experimental discovery of closed cage nearly spherical molecule $C_{60}$ in 1985 by Kroto et al.[2] resulted in the award of Nobel Prize in Chemistry in 1996. The authors also observed a strong resemblance between the family of molecules produced in their experiment with the geodesic domes which were designed and built by R. Buckminster Fuller[3], thus they named those molecules as "fullerene". Moreover, $C_{60}$ molecule was specifically named as "buckminsterfullerene" or simply "buckyball". Thereafter $C_{60}$ and related fullerene molecules becomes the centre of attention due to their exclusive structure and properties[4]. The structure of $C_{60}$ is a kind of polygon termed as truncated icosahedron, consisting of 60 vertices and 32 faces, 12 of which are pentagons and 20 hexagons. The 60 carbon atoms of $C_{60}$ are located at the vertices of the truncated icosahedron and 90 covalent bonds exist between them. However, the length of single bond is larger than double bond of $C_{60}$[5,6] thus $C_{60}$ is not an exact "regular" truncated icosahedrons (Fig. 2). The topologies of $C_{60}$ structure viz. $C_{20}$, $C_{70}$, $C_{80}$, $C_{140}$, $C_{260}$ are also reported in the literature[7,8,9].

---

[1] A. K. Geim & K. S. Novoselov, 'The Rise of Graphene', *Nature Materials* 6 (2007): 183–91, https://doi.org/10.1038/nmat1849.
[2] H. W. Kroto et al., 'C60: Buckminsterfullerene', *Nature* 318, no. 6042 (1985): 162–63, https://doi.org/10.1038/318162a0.
[3] R. B. Fuller et al., "The Dymaxion World of Buckminster Fuller", Anchor Press, Doubleday & Company, Inc., Garden City, New York. c1960.

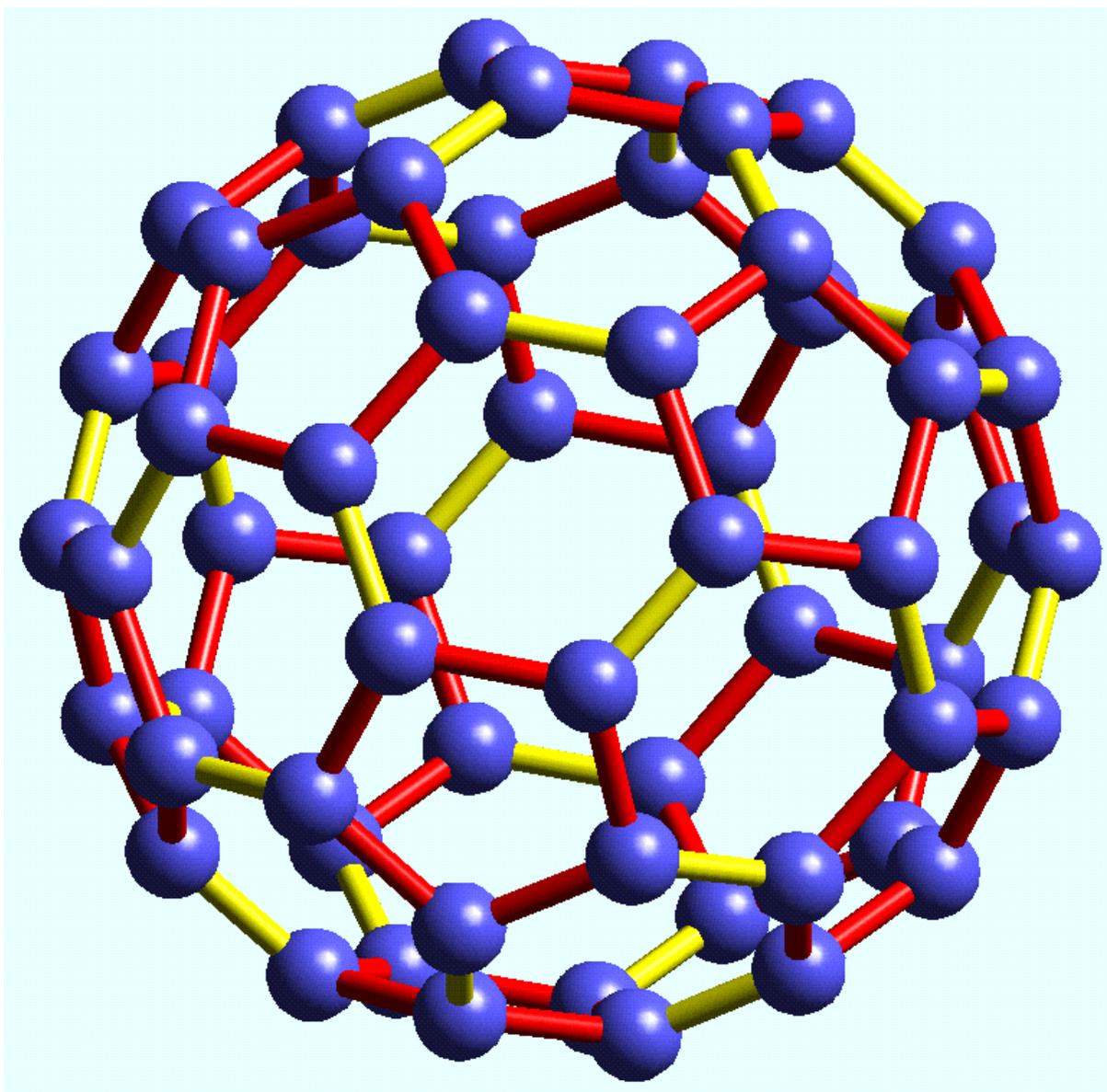

*Fig. 2. Buckyball with single- (red) and double- (yellow) bonds highlighted[10].*

One of the attractive properties of the fullerenes, is the possibility to utilize them as robust containers for other materials. When a fullerene incorporates some material within its carbon cage then it is termed as "endohedral fullerene" (Fig 3). The term "endohedral" was introduced in 1991 by Cioslowski1 et al.[11] and independently by Weiske et al.[12]. If a metallic species is incorporated in fullerene structure then it is called "endohedral metallofullerene"[13] and for incorporation of non metallic species "non-metal doped fullerene" term is used. Due to the novel properties of fullerene[14] it is widely used in different sectors of electronic industry however the fullerene applications in fuel cells and Organic Photovoltaics/Solar Cells will be discussed here.

---

[10] http://www.godunov.com/Bucky/buckyball-3.gif (Accessed on 6 March, 2018)
[11] Jerzy Cioslowski and Eugene D. Fleischmann, 'Endohedral Complexes: Atoms and Ions inside the C60cage', *The Journal of Chemical Physics* 94, no. 5 (1991): 3730–34, https://doi.org/10.1063/1.459744.
[12] Thomas Weiske, 'Endohedral Cluster Compounds: Inclusion of Helium within C60 and C70 through Collision Experiments', *Angewandte Chemie - International Edition* 30, no. 7 (1991): 884–86, https://doi.org/10.1002/anie.199108841.

**2.1 Fullerene applications in fuel cells (FC)**

Consumptions of fossil fuels increases in a rapid manner to cope up with the rising demand for energy and thereby results in high amount of pollution. Scientists are in search of a potential replacement such as renewable energy with high yield and low environmental impact. Under this scenario, FCs appear to be the potential substitutes to be employed as power source in electronic systems (Fig. 3). FC is a kind of electrochemical devices which translate chemical energy into electrical energy[15]. Fullerene derivatives are extensively used in the anode oxidation reaction of direct methanol fuel cells (DMFCs) to augment the activity of methanol electro-oxidation in anode. In 2004 for the first time fullerene was used as a catalyst for methanol oxidation[16] where $C_{60}$ served as a linker system to attach and immobilize Pt nanoparticles on a gold electrode. Monodisperse Pt/$C_{60}$ and PtRu/$C_{60}$ hybrid nanoparticles were produced by Lee et al.[17] employing the reduction of metal precursors in presence of $C_{60}$; a major improvement of methanol oxidation activity was achieved over commercial E-TEK catalysts by the hybrid nanoparticles. Mesoporous fullerene with Pt[18] or Pt/Ru/Sn/W[19] support was also used as catalyst to study anodic performance e.g. methanol oxidation. To study the methanol oxidation reaction; Fullerene was used to coat optically transparent electrodes[20] specially indium tin oxide (ITO)[21] and even glassy carbon (GC)[22] electrode. Moreover the $C_{60}$ is also used in direct formic acid fuel cells (DFAFCs) for the oxidation of formic acid[23] and in direct ethanol fuel cells (DEFCs)[24] for the oxidation of ethanol. In FCs, energy conversion is achieved via oxygen reduction reaction (ORR) which uses pricey and barely accessible platinum metal that necessitates the search for an alternative material. Gir´on et al.[25] synthesized different kinds of fullerene hybrids and metal-free fullerene derivatives which could efficiently catalyze ORRs. Moreover, efficiency of the metal-free fullerenes is as good as metallofullerenes bearing noble metals. Gao et al.[26] prepared nitrogen doped fullerene-like carbon shell by gingko leaves under nitrogen atmosphere at 800°C, which can catalyze ORRs with higher electro-catalytic activity than that of platinum-based electrodes. Fullerene derivatives are also employed as proton conducting membranes. Fullerene composite proton conducting membranes was developed by Tasaki et al.[27] for polymer electrolyte fuel cells operating in low humid situations[28] employing the high electron affinity of fullerenes (2.65 eV for $C_{60}$)[29]. Postnov et al.[30] used impedance spectroscopy to study the nafion-based polyelectrolytes containing $C_{60}$ and its water-soluble derivatives tris-malonate-$C_{60}$ and fullerenol-$C_{60}$ and established that these dopants were capable to substantially enhance proton conductivity of the fullerene composites under low relative humidity. Gasa et al. compared the proton conductivity of different $C_{60}$ derivatives. The studied fullerene derivatives follows the (polyhydroxyl sulphated fullerene:PHSF) > (fullerene phosphonic acid:FPA) > (methanofullerene phosphonic acid: MFPA) > (polyhydroxyfullerenes: PHF) regarding proton conductivity at 20°C and 25% of RH. Thus the PHF was used to synthesize a composite with sulfonated poly (ether ether ketone) (SPEEK) which increased the conductivity and reduced the water solubility compared to SPEEK[31].

---

[13] Hisanori Shinohara, 'Endohedral Metallofullerenes', *Reports on Progress in Physics* 63 (2000): 843–892, https://doi.org/10.1088/0034-4885/63/6/201.
[14] https://pubchem.ncbi.nlm.nih.gov/compound/Buckminsterfullerene
[15] J. M. Andújar and F. Segura, 'Fuel Cells: History and Updating. A Walk along Two Centuries', *Renewable and Sustainable Energy Reviews* 13, no. 9 (2009): 2309–22, https://doi.org/10.1016/j.rser.2009.03.015.
[16] C Roth et al., 'Fullerene-Linked Pt Nanoparticle Assemblies.', *Chemical Communications (Cambridge, England)*, no. May (2004): 1532–33, https://doi.org/10.1039/b404257c.
[17] Gaehang Lee et al., 'Monodisperse Pt and PtRu/C60 Hybrid Nanoparticles for Fuel Cell Anode Catalysts', *Chemical Communications*, no. 33 (2009): 5036–5038, https://doi.org/10.1039/b911068b.
[18] Sujit K. Mondal, 'Synthesis of Mesoporous Fullerene and Its Platinum Composite: A Catalyst for PEMFc', *Journal of The Electrochemical Society* 159, no. 5 (2012): K156–60, https://doi.org/10.1149/2.003206jes.
[19] Karimi Mohammad et al., 'Electrocatalytic Performance of Pt/Ru/Sn/W Fullerene Electrode for Methanol Oxidation in Direct Methanol Fuel Cell', *Journal of Fuel Chemistry and Technology* 41, no. 1 (2013): 91–95, https://doi.org/10.1016/S1872-5813(13)60011-0.
[20] K. Vinodgopal et al., 'Fullerene-Based Carbon Nanostructures for Methanol Oxidation', *Nano Letters* 4, no. 3 (2004): 415–18, https://doi.org/10.1021/nl035028y.
[21] Jianhe Guo et al., 'Single-Crystalline C 60 Crossing Microplates : Preparation , Characterization , and Application as Catalyst Supports for Methanol Oxidation', *Fullerenes, Nanotubes and Carbon Nanostructures* 23 (2014): 424–30, https://doi.org/10.1080/1536383X.2013.843168.
[22] Xuan Zhang and Li Xia Ma, 'Electrochemical Fabrication of Platinum Nanoflakes on Fulleropyrrolidine Nanosheets and Their Enhanced Electrocatalytic Activity and Stability for Methanol Oxidation Reaction', *Journal of Power Sources* 286 (2015): 400–405, https://doi.org/10.1016/j.jpowsour.2015.03.175.

---

[23] Zhengyu Bai, 'A Facile Preparation of Palladium Nanoclusters Supported on Hydroxypropyl-β-Cyclodextrin Modified Fullerene [60] for Formic Acid Oxidation', *International Journal of ELECTROCHEMICAL SCIENCE* 18, no. 7 (2013): 10068–79.
[24] Hamid Reza Barzegar et al., 'Palladium Nanocrystals Supported on Photo-Transformed C60nanorods: Effect of Crystal Morphology and Electron Mobility on the Electrocatalytic Activity towards Ethanol Oxidation', *Carbon* 73 (2014): 34–40, https://doi.org/10.1016/j.carbon.2014.02.028.
[25] Rosa María Girón et al., 'Synthesis of Modified Fullerenes for Oxygen Reduction Reactions', *Journal of Materials Chemistry A* 4, no. 37 (2016): 14284–90, https://doi.org/10.1039/C6TA06573B.
[26] Shuyan Gao et al., 'Nitrogen-Doped Carbon Shell Structure Derived from Natural Leaves as a Potential Catalyst for Oxygen Reduction Reaction', *Nano Energy* 13 (2015): 518–26, https://doi.org/10.1016/j.nanoen.2015.02.031.
[27] Ken Tasaki et al., 'Fullerene Composite Proton Conducting Membranes for Polymer Electrolyte Fuel Cells Operating under Low Humidity Conditions', *Journal of Membrane Science* 281, no. 1–2 (2006): 570–80, https://doi.org/10.1016/j.memsci.2006.04.052.
[28] P Pugazhendhi et al., 'Novel Proton Conducting Fullerene Electrolyte', *ECS Transactions* 12, no. 1 (2008): 21–27, https://doi.org/10.1149/1.2921529.
[29] R S Ruoff et al., 'The Relationship Between the Electron-Affinities and Half-Wave Reduction Potentials of Fullerenes, Aromatic-Hydrocarbons, and Metal-Complexes', *Journal of Physical Chemistry* 99, no. 21 (1995): 8843–50, https://doi.org/10.1021/j100021a060.
[30] D. V. Postnov et al., 'Nafion-Based Composite Solid Electrolytes Containing Water-Soluble Fullerene C60 Derivatives', *Russian Journal of General Chemistry* 86, no. 4 (2016): 890–93, https://doi.org/10.1134/S1070363216040216.

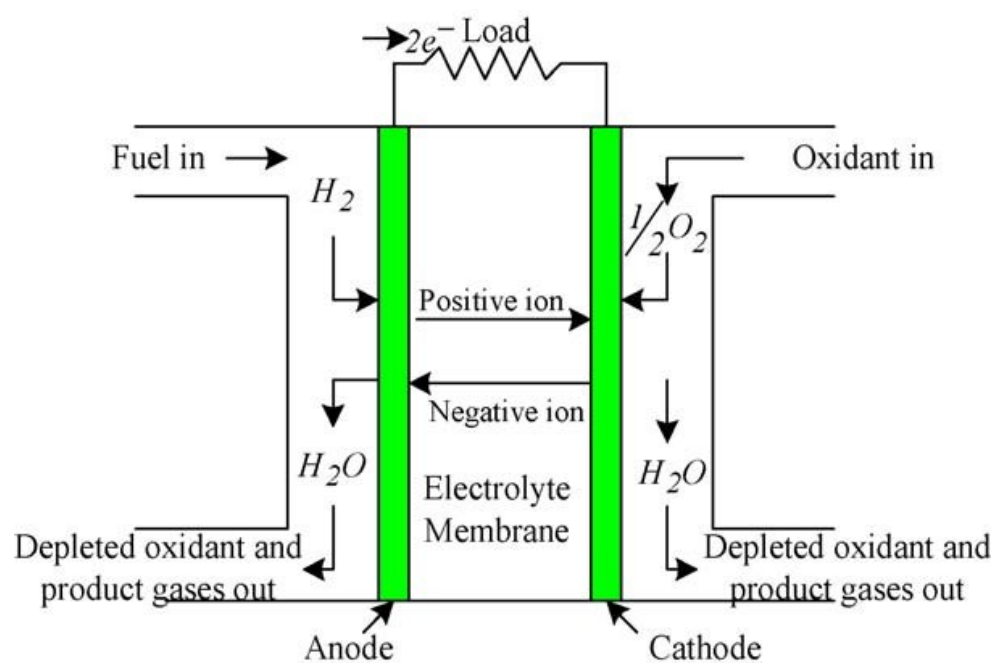

Fig. *3. Fuel cell operation diagram*[32].*(Reprinted with permission from Elsevier)*

**2.2 Fullerene applications in Organic Photovoltaics/Solar Cells**

To cope up with the current energy demand use of clean and green renewable energy such as solar energy is preferable to avoid the depletion of fossil energy resources along with the global warming effect resulted from the greenhouse gas emission. Solar cell[33] is the device that converts the solar radiation into current and fullerene plays major role in enhancing the overall efficiency of the solar cell (Fig. 4).

Bulk-heterojunction polymer solar cells (BHJ-PSCs) made using the composite of semiconducting polymers and different fullerene derivatives, such as [6,6]-phenyl-$C_{71}$-butyric acidmethylester (PC$_{71}$BM) or [6,6]-phenyl-$C_{61}$-butyric acid methylester (PC$_{61}$BM) as the photoactive layer, made ground-breaking expansion in the last decade[34,35]. The demonstrations of efficient electron transfer from polymers to buckminsterfullerene[36] and significant enhancement in power conversion efficiency (PCE) by employing a bulk donor–acceptor heterojunction[37] initiated the research in this field. For single junction bulk-heterojunction organic solar cells (BHJ-OSCs), Chen et al.[38] synthesised a PCE of ~10%, obtained by employing the PTB7-Th polymer with $PC_{71}BM$ acceptor. Also for small molecule based single-junction BHJ-OSCs with $PC_{71}BM$[39], PCE of ~10% is reported. Li et al.[40] achieved 15 % PCE by employing PCBMs in tandem solar cell. He et al.[41] used $C_{70}$-based acceptors to develop another two high-LUMO $C_{70}$ acceptors to achieve PCE of 6.88%.

---

[31] and Ken Tasaki Jeffrey Gasa, Hengbin Wang, Ryan DeSousa, 'Fundamental Characterization of Fullerenes and Their Applications for Proton- Conducting Materials in PEMFC', *ECS Transactions* 11, no. 1 (2007): 131–41, https://doi.org/10.1149/1.2780923.
[32] A. Kirubakaran, Shailendra Jain, and R. K. Nema, 'A Review on Fuel Cell Technologies and Power Electronic Interface', *Renewable and Sustainable Energy Reviews* 13, no. 9 (2009): 2430–40, https://doi.org/10.1016/j.rser.2009.04.004.
[33] Ramasamy Ganesamoorthy, Govindasamy Sathiyan, and Pachagounder Sakthivel, 'Review: Fullerene Based Acceptors for Efficient Bulk Heterojunction Organic Solar Cell Applications', *Solar Energy Materials and Solar Cells* 161, no. August 2016 (2017): 102–48, https://doi.org/10.1016/j.solmat.2016.11.024.

---

[34] Luyao Lu et al., 'Recent Advances in Bulk Heterojunction Polymer Solar Cells', *Chemical Reviews* 115, no. 23 (2015): 12666–731, https://doi.org/10.1021/acs.chemrev.5b00098.

[35] Letian Dou et al., '25th Anniversary Article: A Decade of Organic/Polymeric Photovoltaic Research', *Advanced Materials* 25, no. 46 (2013): 6642–71, https://doi.org/10.1002/adma.201302563.

[36] NS S. Sariciftci et al., 'Photoinduced Electron Transfer from a Conducting Polymer to Buckminsterfullerene.', *Science* 258, no. 5087 (1992): 1474–76, https://doi.org/10.1126/science.258.5087.1474.

[37] G. Yu et al., 'Polymer Photovoltaic Cells: Enhanced Efficiencies via a Network of Internal Donor-Acceptor Heterojunctions', *Science* 270 (1995): 1789–91, https://doi.org/10.1126/science.270.5243.1789.

[38] Jing De Chen et al., 'Single-Junction Polymer Solar Cells Exceeding 10% Power Conversion Efficiency', *Advanced Materials* 27, no. 6 (2015): 1035–41, https://doi.org/10.1002/adma.201404535.

[39] Yongsheng Liu, 'Solution-Processed Small-Molecule Solar Cells: Breaking the 10% Power Conversion Efficiency', *Scientific Reports* 3 (2013): 1–8, https://doi.org/10.1038/srep03356.

[40] Ning Li et al., 'Towards 15% Energy Conversion Efficiency: A Systematic Study of the Solution-Processed Organic Tandem Solar Cells Based on Commercially Available Materials', *Energy & Environmental Science* 6, no. 12 (2013): 3407–13, https://doi.org/10.1039/c3ee42307g.

[41] Dan He et al., 'A Highly Efficient Fullerene Acceptor for Polymer Solar Cells', *Physical Chemistry Chemical Physics* 16 (2014): 7205--7208, https://doi.org/10.1039/c4cp00268g.

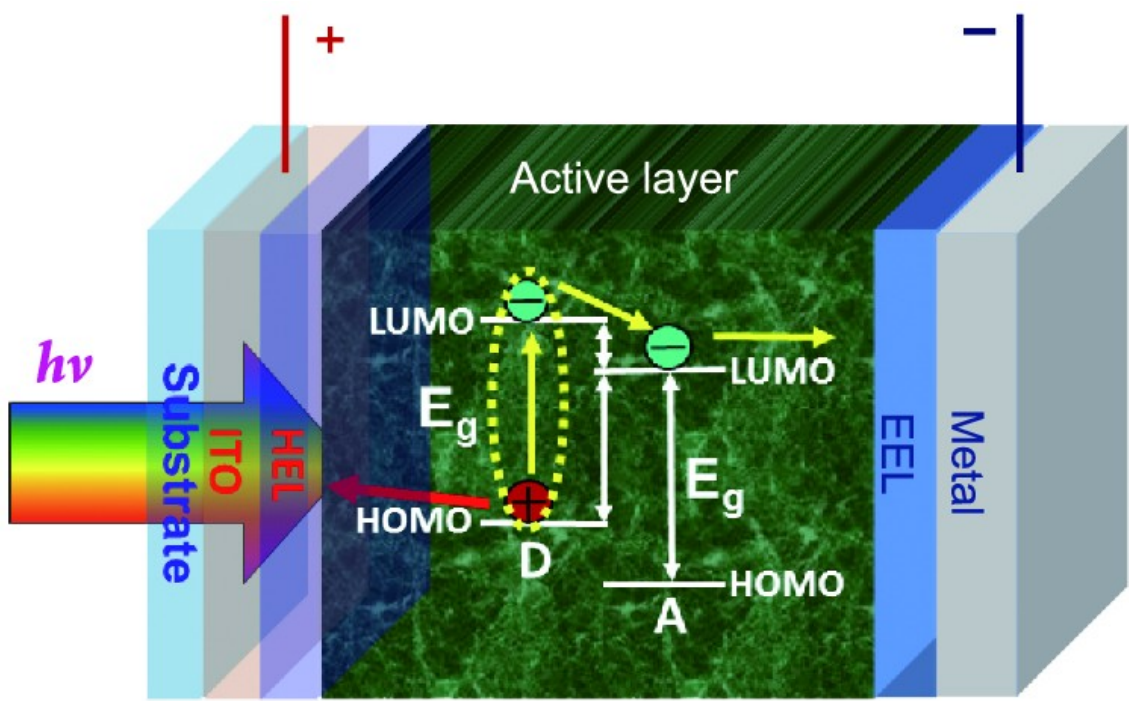

*Fig. 4. Schematic diagram of a polymer solar cell (PSC). The active layer is a mixture of polymer electron donor (D) and fullerene derivative electron acceptor (A)[42]. (Reprinted with permission from Oxford University Press)*

Subsequently, due to the fascinating properties of fullerenes and similar energy levels with perovskites, fullerene derivatives are extensively used in the perovskite light-absorbing layer of PSCs, performing as trap state passivators, interfacial modification materials, or electron transport materials (Fig. 5). Lin et al.[43] employed Indene-$C_{60}$ bisadduct (ICBA) -tran3 to fabricate wide-bandgap (WBG) perovskite solar cell by matching the charge transport layers which resulted in enhanced open circuit voltage ($V_{OC}$) by 60 mV and a stabilized PCE of 18.5%. Bai et al.[44] addressed the issue of water and moisture related issues of perovskite solar cell. To construct highly water-resistant fullerene layer they bonded crosslinkable silane molecules having hydrophobic functional groups with fullerene and also incorporated methylammonium iodide within fullerene layer to achieve n-doping, resulting in perovskite devices with an efficiency of 19.5%. To decrease the interface barrier between electron

---

transport layer (ETL) and metal electrode. Xie et al.[45] used [6, 6]-phenyl-$C_{61}$-butyric acid 2-((2-(dimethylamino) ethyl) (methyl) amino) ethyl ester (PCBDAN) as an interlayer between ETL and metal electrode. The resulted device showed PCE >17.2% with small hysteresis. Kegelmann et al.[46] examined the performance of different electron transport materials in planar n-i-p devices and showed that using $TiO_2$-$PC_{61}BM$ a PCE of 18.4% can be obtained while perovskite solar cell fabricated by Ke et al.[47] showed a PCE of 19.12% employing a PCBM-passivated $SnO_2$ electron selective layer. Yoon et al.[48] reported a hysteresis-free planar $CH_3NH_3PbI_3$ perovskite solar cell with a PCE of 19.1% employing a room-temperature vacuum-processed $C_{60}$ ETL.

For solution processing of metal halide perovskite films, a fullerene derivative (α-bis-PCBM) was employed by Zhang et al.[49] as a templating agent. The α-bis-PCBM covered the grain boundaries and vacancies of the perovskite film, restricted the entry of moisture and passivated voids created in the hole-transporting layer thus increased the perovskites crystallinity, to obtain the PCE of 20.8%.

---

[45] Jiangsheng Xie et al., 'Improved Performance and Air Stability of Planar Perovskite Solar Cells via Interfacial Engineering Using a Fullerene Amine Interlayer', *Nano Energy* 28 (2016): 330–37, https://doi.org/10.1016/j.nanoen.2016.08.048.

[46] Lukas Kegelmann et al., 'It Takes Two to Tango—Double-Layer Selective Contacts in Perovskite Solar Cells for Improved Device Performance and Reduced Hysteresis', *ACS Applied Materials & Interfaces* 9, no. 20 (2017): 17245–55, https://doi.org/10.1021/acsami.7b00900.

[47] Weijun Ke et al., 'Cooperative Tin Oxide Fullerene Electron Selective Layers for High-Performance Planar Perovskite Solar Cells', *J. Mater. Chem. A* 4, no. 37 (2016): 14276–83, https://doi.org/10.1039/C6TA05095F.

[48] Heetae Yoon et al., 'Hysteresis-Free Low-Temperature-Processed Planar Perovskite Solar Cells with 19.1% Efficiency', *Energy Environ. Sci.* 9, no. 7 (2016): 2262–66, https://doi.org/10.1039/C6EE01037G.

[49] Fei Zhang et al., 'Isomer-Pure Bis-PCBM-Assisted Crystal Engineering of Perovskite Solar Cells Showing Excellent Efficiency and Stability', *Advanced Materials* 29, no. 17 (2017): 1606806 (1-7), https://doi.org/10.1002/adma.201606806.

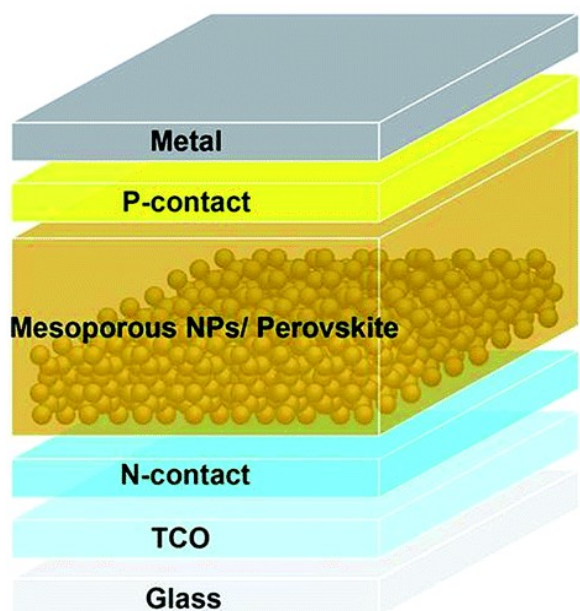 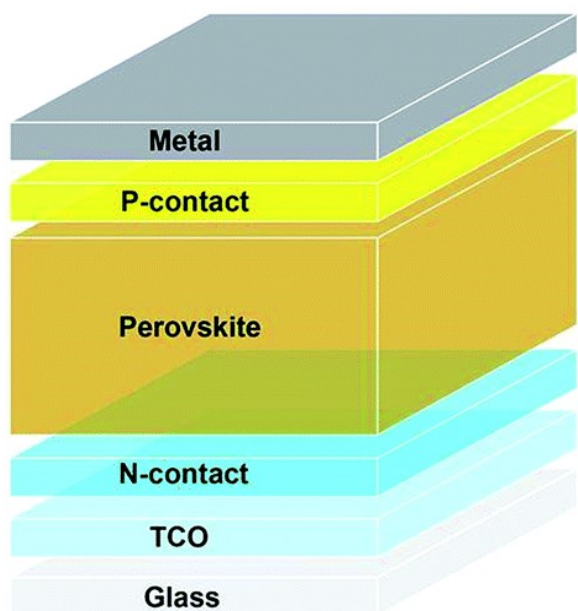
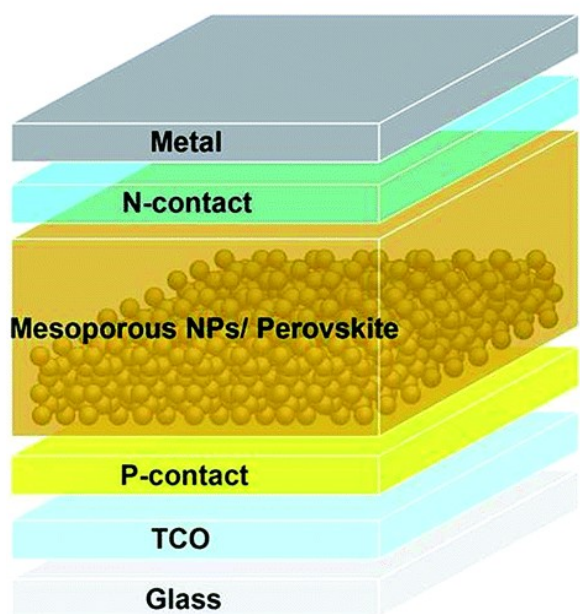 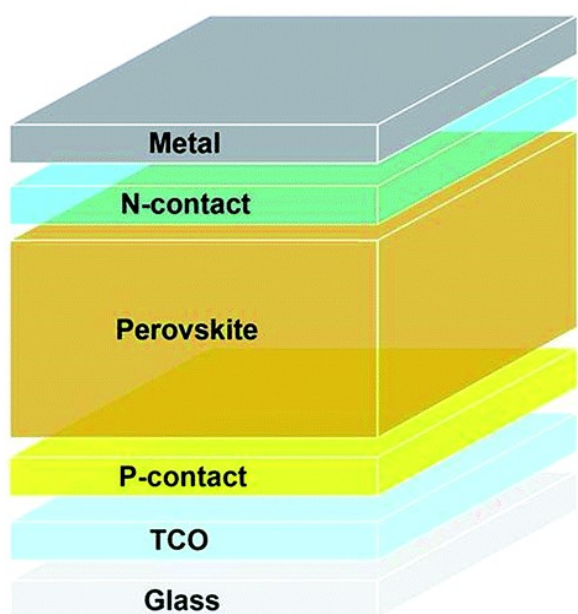

*Fig. 5. Device structures of (a) n-i-p mesoscopic, (b) n-i-p planar, (c) p-i-n mesoscopic and (d) p-i-n planar perovskite solar cells[50].(Reprinted with permission from Royal Society of Chemistry)*

---

[50] Ming-Hsien Li et al., 'Inorganic P-Type Contact Materials for Perovskite-Based Solar Cells', *J. Mater. Chem. A* 3, no. 17 (2015): 9011–19, https://doi.org/10.1039/C4TA06425A.

Endohedral Fullerenes are also used as acceptors in PSCs. Ross et al.[51] used trimetallic nitride endohedral fullerenes (TNEFs), $Lu_3N@C_{80}$ to be precise, as acceptor materials to be used in photovoltaic devices that have an 890mV open circuit voltage with PCE > 4%. The same group also demonstrated that the organic photovoltaic device (OPV) made with $Lu_3N@C_{80}$ has higher open circuit voltage in comparison to empty cage fullerene acceptors[52].

## 3. Carbon nanotube

.The tubular carbon nanostructures were first detected in 1952 by Radushkevich and Lukyanovich[53]. Later on, Oberlin et al.[54] reported the presence of hollow carbon fibres having diameters in nanometer regime. Nearly after two decades, Iijima[55] published the clear images of carbon nanotubes (CNTs) in Nature journal which created that global interest (Fig. 6).

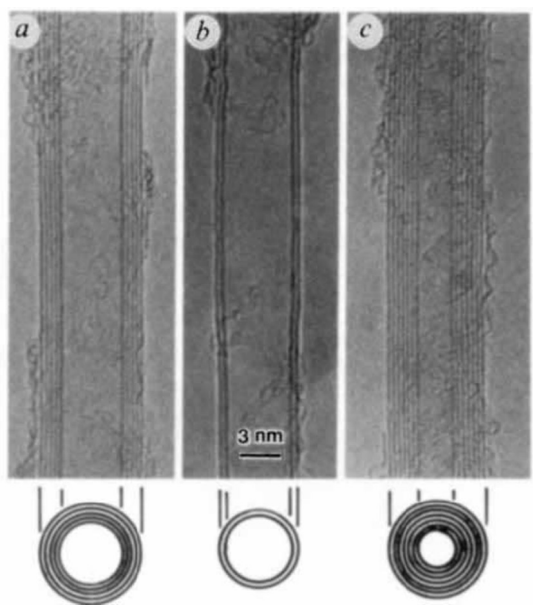

*Fig 6. Electron micrographs of microtubules of graphitic carbon. (Reprinted with permission from Springer Nature)*

---

[51] Russel B Ross et al., 'Endohedral Fullerenes for Organic Photovoltaic Devices', *Nature Materials* 8, no. 3 (2009): 208–12, https://doi.org/10.1038/nmat2379.
[52] Russel B. Ross et al., 'Tuning Conversion Efficiency in Metallo Endohedral Fullerene-Based Organic Photovoltaic Devices', *Advanced Functional Materials* 19, no. 14 (2009): 2332–37, https://doi.org/10.1002/adfm.200900214.
[53] L. V. Radushkevich, 'The Structure of Carbon Forming in Thermal Decomposition of Carbon Monoxide on an Iron Catalyst', *Soviet Journal of Physical Chemistry*, 1952.
[54] A.Oberlin. M.Endo .T.Koyama, 'Filamentous Growth of Carbon through Benzene Decomposition', *Journal of Crystal Growth* 32, no. 3 (1976): 335–49, https://doi.org/10.1016/0022-0248(76)90115-9.
[55] SUMIO IIJIMA, 'Helical Microtubules of Graphitic Carbon', *Nature* 354, no. 07 Novembe (1991): 56–58, https://doi.org/10.1038/354056a0.

Nanotubes are almost one dimensional in structure with high aspect ratio and display a unique combination of novel properties[56,57,58,59]. Three different types of nanotube can be formed by rolling the graphene sheet along different directions: armchair, zigzag, and chiral. Armchair CNTs are metallic. Zigzag and chiral nanotubes can be semimetals or semiconductors. The term zigzag and armchair refer to the arrangement of hexagons around the perimeter of a CNT. CNTs can also be envisaged as sheets of carbon atoms rolled up into tubes and they can consist of one (single-wall) or more (multi-wall) graphitic layers. Depending on the roll-up direction SWCNTs can have different structures namely zigzag, armchair or chiral (Fig. 7).

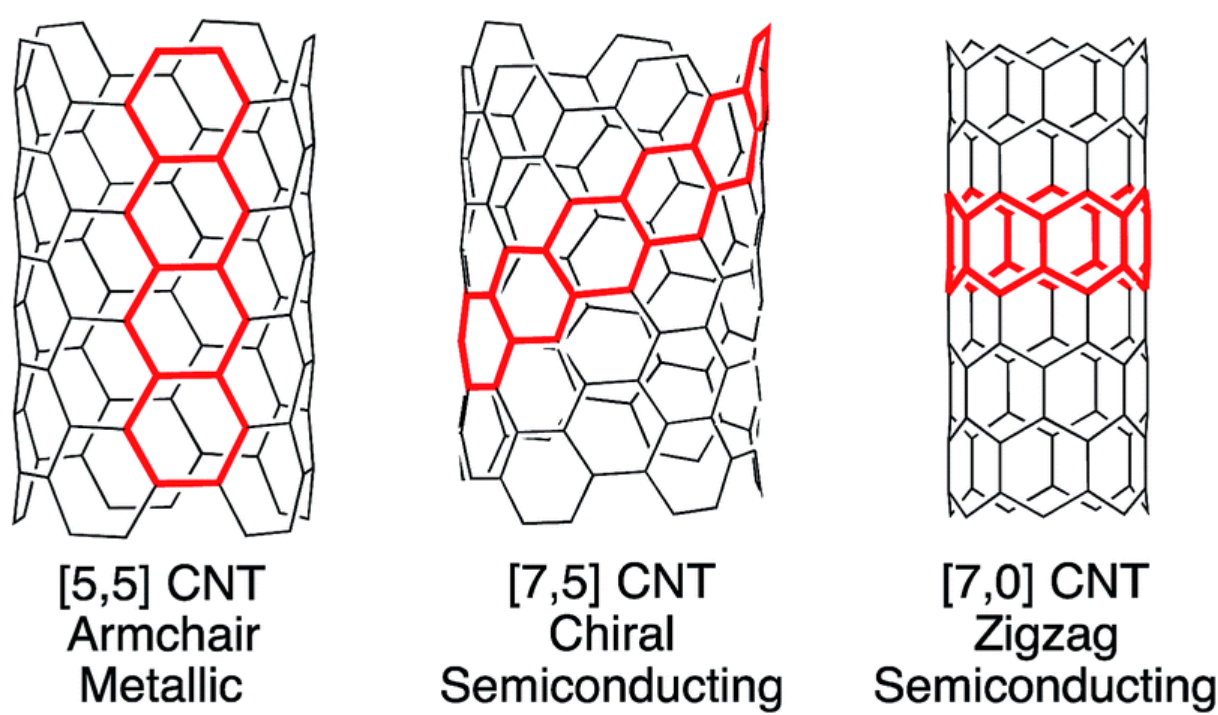

*Fig. 7. Schematics of various types of single-walled carbon nanotubes[60]. (Reprinted with permission from Royal Society of Chemistry)*

---

[56] P. Chen, 'High H2 Uptake by Alkali-Doped Carbon Nanotubes Under Ambient Pressure and Moderate Temperatures', *Science* 285, no. 5424 (1999): 91–93, https://doi.org/10.1126/science.285.5424.91.

[57] Hongjie Dai et al., 'Nanotubes as Nanoprobes in Scanning Probe Microscopy', *Nature* 384, no. 6605 (1996): 147–50, https://doi.org/10.1038/384147a0.

[58] Koichi Hata Yahachi Saito, Koji Hamaguchi, 'Conical Beams from Open Nanotubes', *Nature* 389, no. 6651 (1997): 554–55, https://doi.org/10.1038/39221.

[59] Sander J Tans, Alwin R M Verschueren, and Cees Dekker, 'Room-Temperature Transistor Based on a Single Carbon Nanotube', *Nature* 393, no. 6680 (1998): 669–72, https://doi.org/10.1038/29954.

[60] Thomas J. Sisto et al., 'Towards Pi-Extended Cycloparaphenylenes as Seeds for CNT Growth: Investigating Strain Relieving Ring-Openings and Rearrangements', *Chem. Sci.* 7, no. 6 (2016): 3681–88, https://doi.org/10.1039/C5SC04218F.

The structure of multi-walled nanotube can be described based on two models– the "Parchment model" and the "Russian doll" model (Fig. 8). In the "Parchment model", a single graphene sheet is rolled in around itself multiple times to form a spiral, like a roll of paper or parchment whereas in the "Russian doll" model, sheets of graphite are organized in concentric cylinders. However, there are reports on the structure of MWCNTs which indicate that the structure of MWCNTs may be a defective roll[61,62] also called *"Papier mâché"* or a mixture of "Russian doll" and "Swiss roll" structures[63].

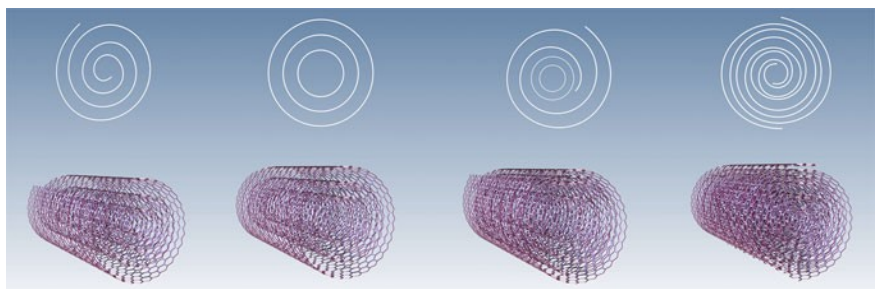

*Fig 8. Schematics of various types of multi-walled carbon nanotubes,[64] (Reprinted with permission from Springer Nature)*

Among different nanomaterials CNT is a very interesting one as it provides an extremely small inner hollow core, nearly one-dimensional space, for storage of materials. Thus a little engineering by filling the core of CNT with required materials reveals an extraordinary structure that encompasses the properties of both the host CNT and the guest filling material. Thus whenever the core of the CNT is filled with any material then a new term "filled CNT"[65] is evolved.

---

[61] O. Zhou et al., 'Defects in Carbon Nanostructures', *Science* 263, no. 5154 (1994): 1744–47, https://doi.org/10.1126/science.263.5154.1744.
[62] Irene Suarez-Martinez et al., 'Dislocations in Carbon Nanotube Walls.', *Journal of Nanoscience and Nanotechnology* 7, no. 10 (2007): 3417–20, https://doi.org/10.1166/jnn.2007.844.
[63] Yutaka Maniwa et al., 'Multiwalled Carbon Nanotubes Grown in Hydrogen Atmosphere: An x-Ray Diffraction Study', *Physical Review B* 64, no. 7 (2001): 073105 (1-4), https://doi.org/10.1103/PhysRevB.64.073105.
[64] M H M O Hamanaka, V P Mammana, and P J Tatsch, 'Review of Field Emission from Carbon Nanotubes: Highlighting Measuring Energy Spread', in *NanoCarbon 2011. Carbon Nanostructures*, ed. Avellaneda C., vol. 3 (Springer, Berlin, Heidelberg, 2013), 1–32, https://doi.org/10.1007/978-3-642-31960-0_1.
[65] Joydip Sengupta et al., 'Lithographically Defined Site-Selective Growth of Fe Filled Multi-Walled Carbon Nanotubes Using a Modified Photoresist', *Carbon* 48, no. 8 (2010): 2371–75, https://doi.org/10.1016/j.carbon.2010.02.036.

There are numerous potential application sectors of CNT such as flat panel display[66,67], sensor[68], wearable electronics[69], solar cell[70], Li ion battery[71], stimuli-responsive material[72] and even in textile industry[73]. Here we will discuss the use of CNT in wearable electronics and stimuli-responsive material.

### 3.1 Carbon nanotube application in wearable electronics

The speedy development of portable electronics necessitates the advancement of flexible energy conversion and/or storage devices with superior performance[74]. Due to the exclusive structure of wearable electronics it shows sufficient flexibility to be easily incorporated into any form of textiles by employing conventional woven technology. The superior mechanical strength, high electrical conductivity, good flexibility along with supreme electrochemical property makes CNT a perfect fiber-shaped electrode[75] for wearable electronics[76]. In order to achieve high quality CNT fiber the principle criterion is to ensure that CNTs are aligned along the axial direction of a fiber. There are various approaches to synthesize CNT fibbers in the optimized way viz. spinning from a CNT film[77], dry spinning from a vertically aligned

---

[66] Jongmin Kim et al., 'Flat Panel Display Using Carbon Nanotube', in *2006 International Symposium on Discharges and Electrical Insulation in Vacuum* (IEEE, 2006), 888–89, https://doi.org/10.1109/DEIV.2006.357448.

[67] A. A. Talin et al., 'Development of Field Emission Flat Panel Displays at Motorola', *MRS Proceedings* 508 (10 January 1998): 175–77, https://doi.org/10.1557/PROC-508-175.

[68] Irina V. Zaporotskova et al., 'Carbon Nanotubes: Sensor Properties. A Review', *Modern Electronic Materials* 2, no. 4 (2016): 95–105, https://doi.org/10.1016/j.moem.2017.02.002.

[69] Jeong Hun Kim et al., 'Simple and Cost-Effective Method of Highly Conductive and Elastic Carbon Nanotube/Polydimethylsiloxane Composite for Wearable Electronics', *Scientific Reports* 8, no. 1 (2018): 1375 (1-11), https://doi.org/10.1038/s41598-017-18209-w.

[70] Utkarsh Kumar et al., 'Carbon Nanotube: Synthesis and Application in Solar Cell', *Journal of Inorganic and Organometallic Polymers and Materials* 26, no. 6 (2016): 1231–42, https://doi.org/10.1007/s10904-016-0401-z.

[71] Poonam Sehrawat, C. Julien, and S. S. Islam, 'Carbon Nanotubes in Li-Ion Batteries: A Review', *Materials Science and Engineering B: Solid-State Materials for Advanced Technology* 213 (2016): 12–40, https://doi.org/10.1016/j.mseb.2016.06.013.

[72] R. Bafkary and S. Khoee, 'Carbon Nanotube-Based Stimuli-Responsive Nanocarriers for Drug Delivery', *RSC Advances* 6, no. 86 (2016): 82553–65, https://doi.org/10.1039/C6RA12463A.

[73] Sheila Shahidi and Bahareh Moazzenchi, 'Carbon Nanotube and Its Applications in Textile Industry – A Review', *The Journal of The Textile Institute*, 2018, 1–14, https://doi.org/10.1080/00405000.2018.1437114.

[74] Lin Li et al., 'Advances and Challenges for Flexible Energy Storage and Conversion Devices and Systems', *Energy & Environmental Science* 7, no. 7 (2014): 2101–22, https://doi.org/10.1039/c4ee00318g.

[75] Weibang Lu et al., 'State of the Art of Carbon Nanotube Fibers: Opportunities and Challenges', *Advanced Materials* 24, no. 14 (2012): 1805–33, https://doi.org/10.1002/adma.201104672.

[76] Tao Chen et al., 'Novel Solar Cells in a Wire Format', *Chemical Society Reviews* 42, no. 12 (2013): 5031–41, https://doi.org/10.1039/c3cs35465b.

[77] Wenjun Ma et al., 'Monitoring a Micromechanical Process in Macroscale Carbon Nanotube Films and Fibers', *Advanced Materials* 21, no. 5 (2009): 603–8, https://doi.org/10.1002/adma.200801335.

CNT array[78], wet spinning from CNT solution[79] and in-situ spinning from a CNT aerogel directly synthesized in a CVD reactor[80].

Chen et al.[81] employed the semiconducting CNT fiber in dye-sensitized solar cells (DSSCs), in which the highly aligned nanotubes in the fiber permit the charges to separate and transport along the fibers proficiently to achieve high performance. As a result, PCE of 2.2% was obtained using single CNT fiber. Cai et al.[82] synthesised a highly efficient and low-cost solution process to fabricate a novel dye-sensitized photovoltaic wire using CNT and proposed that the PCE could be significantly enhanced via modification of the working electrode, viz. incorporation of $TiO_2$ nanomaterials. Chen et al.[83] used this idea and coated CNT fiber with $TiO_2$ layer by employing a facile dip-coating method to achieve PCE of about 3%. Chen et al.[84] further modified the structure with aligned carbon and $TiO_2$ nanotubes to augment the charge separation and transport in such a way that the resulting wire cell showed a high PCE up to ∼ 4.6%. The efficiency of the fiber-shaped DSSCs can be enhanced up to 4.85% by the introduction of platinum nanoparticles[85] and up to 7.33% by means of an organic thiolate/disulfide redox couple as electrolyte[86]. Qiu et al.[87] fabricated a fiber-shape perovskite solar cell with a core (steel wire) - shell (CNT film) structure to obtain PCE of 3.3%. Yang et al.[88] reported a stretchable, wearable DSSC textile employing spring-like titanium wire as the working electrode and electrically conducting fiber as counter electrode which exhibited PCE of 7.13%. Zhang et al.[89] used a coaxial structure comprised of an aligned CNT sheet anode (outside) and a modified metal wire cathode (inside), with an electroluminescent polymer interlayer, to fabricate a fibre-shaped, weavable polymer light-emitting electrochemical cell (PLEC). The resulted PLEC was flexible, and its brightness was remained at above 90% of its maximum even after bending with a radius of curvature of 6 mm for 100 cycles. Ren et al.[90] fabricated a fiber-shaped electric double-layer capacitor (EDLC) by twisting two aligned MWCNT/ ordered mesoporous carbon (OMC) composite fibers as electrodes. The resultant composite fiber electrode incorporated the structural and property wise advantages of both the components and the EDLC wire depicts long life stability with large specific capacitance. A coaxial EDLC fiber with a high electrochemical performance was fabricated by Chen et al.[91] using the aligned CNT fiber (inside) and CNT sheet (outside), which acts as two electrodes with a polymer gel inserted between them and provides a maximum discharge capacitance of 59 $Fg^{-1}$. Wang et al.[92] synthesised two-ply yarn supercapacitors with capacitance 38 $mFcm^{-2}$ employing CNT yarns and polyaniline nanowires. Lee et al.[93] developed CNT/PEDOT composite fiber supercapacitor using a facile technology called gradient biscrolling, that offers fast- ion-transport yarn comprised of layers of conducting-polymer-infiltrated CNT sheet which are scrolled into almost 20 μm diameter yarn to achieve volumetric capacitances of 147 $Fcm^{-3}$ (liquid electrolyte) and 145 $Fcm^{-3}$ (solid electrolyte). A stable, all-nanotube stretchable supercapacitor was developed by Gilshteyn et al.[94] employing SWCNT film electrodes and a boron nitride nanotube separator. The supercapacitor depicted low equivalent series resistance of 4.6 Ω with specific capacitance value of 82 $Fg^{-1}$.

---

[78] Kaili Jiang, Qunqing Li, and Shoushan Fan, 'Spinning Continuous Carbon Nanotube Yarns', *Nature* 419, no. 6909 (2002): 801, https://doi.org/10.1038/419801a.
[79] B. Vigolo et al., 'Macroscopic Fibers and Ribbons of Oriented Carbon Nanotubes', *Science* 290, no. 5495 (2000): 1331–34, https://doi.org/10.1126/science.290.5495.1331.
[80] Ya Li Li, Ian A. Kinloch, and Alan H. Windle, 'Direct Spinning of Carbon Nanotube Fibers from Chemical Vapor Deposition Synthesis', *Science* 304, no. 5668 (2004): 276–78, https://doi.org/10.1126/science.1094982.
[81] Tao Chen et al., 'Flexible, Light-Weight, Ultrastrong, and Semiconductive Carbon Nanotube Fibers for a Highly Efficient Solar Cell', *Angewandte Chemie - International Edition* 50, no. 8 (2011): 1815–19, https://doi.org/10.1002/anie.201003870.
[82] Fangjing Cai, Tao Chen, and Huisheng Peng, 'All Carbon Nanotube Fiber Electrode-Based Dye-Sensitized Photovoltaic Wire', *Journal of Materials Chemistry* 22, no. 30 (2012): 14856–14860, https://doi.org/10.1039/c2jm32256k.
[83] T Chen et al., 'Intertwined Aligned Carbon Nanotube Fiber Based Dye-Sensitized Solar Cells', *Nano Letters* 12, no. 5 (2012): 2568–72, https://doi.org/10.1021/nl300799d.
[84] Tao Chen et al., 'Designing Aligned Inorganic Nanotubes at the Electrode Interface: Towards Highly Efficient Photovoltaic Wires', *Advanced Materials* 24, no. 34 (2012): 4623–28, https://doi.org/10.1002/adma.201201893.
[85] Sen Zhang et al., 'Porous , Platinum Nanoparticle-Adsorbed Carbon Nanotube Yarns for Efficient Fiber Solar Cells', *ACS Nano* 6, no. 8 (2012): 7191–98, https://doi.org/10.1021/nn3022553.
[86] Shaowu Pan et al., 'Efficient Dye-Sensitized Photovoltaic Wires Based on an Organic Redox Electrolyte', *Journal of the American Chemical Society* 135, no. 29 (2013): 10622–25, https://doi.org/10.1021/ja405012w.
[87] Longbin Qiu et al., 'Integrating Perovskite Solar Cells into a Flexible Fiber', *Angewandte Chemie International Edition* 53, no. 39 (2014): 10425–28, https://doi.org/10.1002/anie.201404973.
[88] Zhibin Yang et al., 'Stretchable, Wearable Dye-Sensitized Solar Cells', *Advanced Materials* 26, no. 17 (2014): 2643–47, https://doi.org/10.1002/adma.201400152.

Yang et al.[95] developed a fiber- shaped highly stretchable, supercapacitor by wrapping two layers of CNT sheets, which act as two electrodes, on an elastic fiber. The supercapacitor depicted a high specific capacitance of 18 F/g after stretch by 75% for 100 cycles. Chen et al.[96] wrapped a CNT film around pre-stretched elastic wires and then twisted two such CNT-wrapped elastic wires to create wire-shaped supercapacitor with a large device capacitance up to 30.7 Fg$^{-1}$ and elasticity up to 350% strain. Zhang et al.[97] developed a fiber-shaped supercapacitor employing two aligned CNT/polyaniline composite sheets as electrodes to achieve specific capacitance of approximately 79.4 Fg$^{-1}$ even after being stretched at a strain of 300% for 5000 cycles. Multifunctional wearable supercapacitors which can be used for self-healing[98], shape-memory[99], electrochromic[100] etc. have also created considerable attention from scientific community.

Ren et al.[101] synthesised a fiber-shaped lithium ion battery (LIB) by twisting an aligned MWCNT/MnO$_2$ composite fiber and Li wire which acted as positive and negative electrodes, respectively. During the charge and discharge process, the achieved energy densities were 92.84 and 35.74 mWh/cm$^3$ with the power densities 3.87 and 2.43 W/cm$^3$, respectively. Lin et al.[102] used the aligned MWCNT/Si composite fiber as a working electrode with a Li wire as both counter and reference electrode to synthesise LIB with specific capacities 1670 mAh/g. Two aligned CNT-lithium manganate and CNT-Si/CNT composite yarns were sequentially wrapped onto a cotton fiber by Weng et al.[103] to fabricate a coaxial fiber battery with an areal energy density of 4.5 mWhcm$^{-2}$ and linear energy density of 0.75 mWh cm$^{-1}$. Fiber-shaped, super-stretchy, LIBs with a strain of 600% were fabricated by Zhang et al.[104] employing a new twisted structure of two highly aligned CNT composite fibers which acted as positive and negative electrodes, which later coated with a thin layer of gel electrolyte. Wearable integrated energy device, the "energy wire", which can store the electric energy produced by an energy conversion device (viz., solar cell) into an energy storage device itself (viz., battery and supercapacitor) has created huge interest in recent past[105]. Chen et al.[106] wrapped aligned CNT fibers around a $TiO_2$ nanowire and treated the ends of the wire-shaped structure with an electrolyte and a light-sensitive dye to create photoelectric-conversion and energy-storage regions in the same device. The "energy wire" revealed a large overall photoelectric conversion and storage efficiency of 1.5 %. Zhang et al.[107] employed aligned MWCNT sheet and titania nanotube-modified Ti wire as two electrodes to fabricate integrated PSC and electrochemical supercapacitor with a stable and flexible fiber format. Pan et al.[108] fabricated a transparent, flexible, thin and lightweight supercapacitor using CNT /polyaniline (PANI) composite fiber and it is further employed to produce a novel energy textile which can translate solar energy to electric energy. The resultant supercapacitor textile depicted high specific capacitance of 272.7 $Fg^{-1}$ and a large photoelectric conversion and high storage efficiency of 2.1% was obtained. Different wearable electronic device fabricated using CNT are depicted in Fig. 9.

---

[96] Tao Chen et al., 'High-Performance, Stretchable, Wire-Shaped Supercapacitors', *Angewandte Chemie - International Edition* 54, no. 2 (2015): 618–22, https://doi.org/10.1002/anie.201409385.
[97] Zhitao Zhang et al., 'Superelastic Supercapacitors with High Performances during Stretching', *Advanced Materials* 27, no. 2 (2015): 356–62, https://doi.org/10.1002/adma.201404573.
[98] Hao Sun et al., 'Self-Healable Electrically Conducting Wires for Wearable Microelectronics', *Angewandte Chemie - International Edition* 53, no. 36 (2014): 9526–31, https://doi.org/10.1002/anie.201405145.
[99] Jue Deng et al., 'A Shape-Memory Supercapacitor Fiber', *Angewandte Chemie - International Edition* 54, no. 51 (2015): 15419–23, https://doi.org/10.1002/anie.201508293.
[100] Xuli Chen et al., 'Electrochromic Fiber-Shaped Supercapacitors', *Advanced Materials* 26, no. 48 (2014): 8126–32, https://doi.org/10.1002/adma.201403243.
[101] Jing Ren et al., 'Twisting Carbon Nanotube Fibers for Both Wire-Shaped Micro-Supercapacitor and Micro-Battery', *Advanced Materials* 25, no. 8 (2013): 1155–59, https://doi.org/10.1002/adma.201203445.
[102] Huijuan Lin et al., 'Twisted Aligned Carbon Nanotube/Silicon Composite Fiber Anode for Flexible Wire-Shaped Lithium-Ion Battery', *Advanced Materials* 26, no. 8 (2014): 1217–22, https://doi.org/10.1002/adma.201304319.
[103] Wei Weng et al., 'Winding Aligned Carbon Nanotube Composite Yarns into Coaxial Fiber Full Batteries with High Performances', *Nano Letters* 14, no. 6 (2014): 3432–38, https://doi.org/10.1021/nl5009647.
[104] Ye Zhang et al., 'Super-Stretchy Lithium-Ion Battery Based on Carbon Nanotube Fiber', *Journal of Materials Chemistry A* 2, no. 29 (2014): 11054–59, https://doi.org/10.1039/c4ta01878h.

---

[105] Tao Chen, Zhibin Yang, and Huisheng Peng, 'Integrated Devices to Realize Energy Conversion and Storage Simultaneously', *ChemPhysChem* 14, no. 9 (2013): 1777–82, https://doi.org/10.1002/cphc.201300032.
[106] Tao Chen et al., 'An Integrated "Energy Wire" for Both Photoelectric Conversion and Energy Storage', *Angewandte Chemie - International Edition* 51, no. 48 (2012): 11977–80, https://doi.org/10.1002/anie.201207023.
[107] Zhitao Zhang et al., 'Integrated Polymer Solar Cell and Electrochemical Supercapacitor in a Flexible and Stable Fiber Format', *Advanced Materials* 26, no. 3 (2014): 466–70, https://doi.org/10.1002/adma.201302951.
[108] Shaowu Pan et al., 'Novel Wearable Energy Devices Based on Aligned Carbon Nanotube Fiber Textiles', *Advanced Energy Materials* 5, no. 4 (2015): 1401438 (1-7), https://doi.org/10.1002/aenm.201401438.

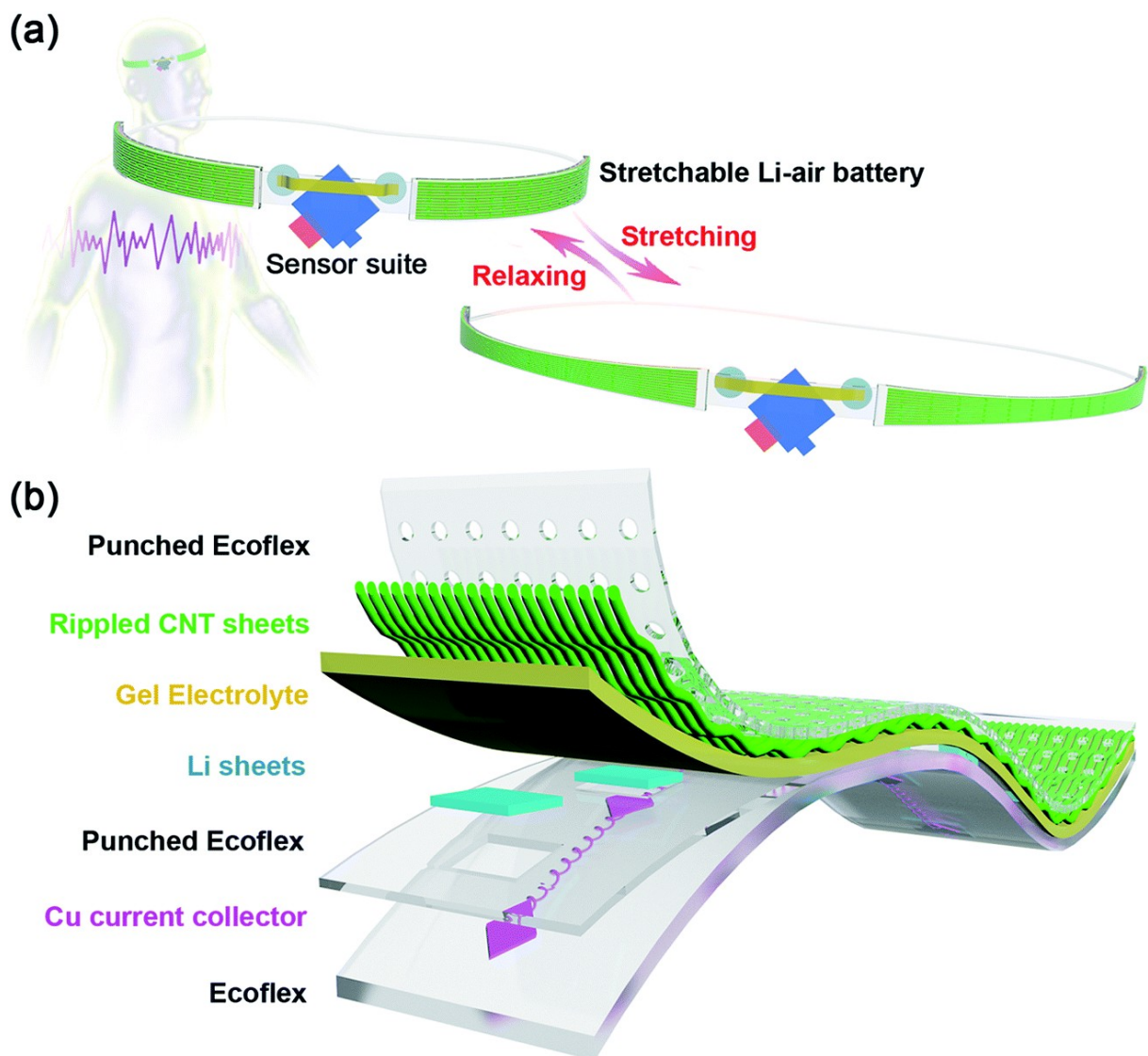

*Fig. 9. a) Potential application of the stretchable Li-air batteries as elastic "energetic straps" to power a sensor suite for wearable physiological monitoring. b) Multi-layered structure of the stretchable Li-air battery[109] (Reprinted with permission from Royal Society of Chemistry)*

**3.2 Carbon nanotube application in display technology**

CNTs are used as the polarizer in liquid crystal displays (LCDs), the field emitter in field emission devices (FEDs), and the transparent electrode in organic light emitting device. CNTs are also employed in a few new kinds of displays, such as thermochromic displays, flexible fiber displays and incandescence displays. The a basic technical requirement for the production of FED is the mass production of uniform CNT emitter over a large area and CVD technique[110] is commonly employed to achieve the goal. Gangloff et al.[111] used lithography for the production of integrated-gate nanocathode array with a low turn-on voltage of 25 V where a nanocathode was composed of a single CNT per gate aperture. Using the field emission characteristics of CNT, Wang et al.[112] fabricated a matrix addressable diode flat panel display employing a CNT-epoxy composite as the electron emission source. Choi et al.[113] developed a fully sealed FED with 4.5 inch in size using SWCNT-organic binders. Saito et al.[114] manufactured cathode ray tubes (CRTs) with adequate luminance, long life and stable electron emission, by employing field emitters that composed of MWCNTs. Wei et al.[115] devised an effective process to fabricate linear, cold cathodes with CNT emitters. Furthermore, luminescent tubes were constructed employing a transparent CNT film as the anode and the CNT wire as the cathode. Wei et al.[116] developed field electron emitters by fragmenting a continuous MWCNT yarn into segments which served as field emitters and resulted emission current can reach several milliamperes with high stability. The same group also used joule heating[117] to construct aligned surface structure using SWCNT bundles and the bundles offer emission currents up to 100 μA, and a pixel was developed for lateral field emission displays. Backlight unit is another application area of the CNT field emission structure in display. Chung et al.[118] constructed row–column matrix-addressable FEDs via CNT emitters which can be scaled up over 40 in. in diagonal on glass substrates with low cost. Choi et al.[119] developed a field emission backlight unit employing a CNT emitter and studied its field emission characteristics like uniformity, stability and emission current.

---

[109] Lie Wang et al., 'Stretchable Lithium-Air Batteries for Wearable Electronics', *Journal of Materials Chemistry A* 4, no. 35 (2016): 13419–24, https://doi.org/10.1039/c6ta05800k.

---

[110] S Fan et al., 'Self-Ooriented Regular Arrays of Carbon Nanotubes and Their Field Emission Properties', *Science* 283, no. 5401 (1999): 512–14, https://doi.org/10.1126/science.283.5401.512.
[111] L. Gangloff et al., 'Self-Aligned, Gated Arrays of Individual Nanotube and Nanowire Emitters', *Nano Letters* 4, no. 9 (2004): 1575–79, https://doi.org/10.1021/nl049401t.
[112] Q. H. Wang et al., 'A Nanotube-Based Field-Emission Flat Panel Display', *Applied Physics Letters* 72, no. 22 (1998): 2912–13, https://doi.org/10.1063/1.121493.
[113] W. B. Choi et al., 'Fully Sealed, High-Brightness Carbon-Nanotube Field-Emission Display', *Applied Physics Letters* 75, no. 20 (1999): 3129–31, https://doi.org/10.1063/1.125253.
[114] Yahachi Saito, Sashiro Uemura, and Koji Hamaguchi, 'Cathode Ray Tube Lighting Elements with Carbon Nanotube Field Emitters', *Japanese Journal of Applied Physics, Part 2: Letters* 37, no. 3 B (1998): L346–L348, https://doi.org/10.1143/JJAP.37.L346.
[115] Yang Wei et al., 'Cold Linear Cathodes with Carbon Nanotube Emitters and Their Application in Luminescent Tubes', *Nanotechnology* 18, no. 32 (2007): 325702 (1-5), https://doi.org/10.1088/0957-4484/18/32/325702.
[116] Yang Wei et al., 'Efficient Fabrication of Field Electron Emitters from the Multiwalled Carbon Nanotube Yarns', *Applied Physics Letters* 89, no. 6 (2006): 063101 (1-3), https://doi.org/10.1063/1.2236465.
[117] Yang Wei et al., 'Breaking Single-Walled Carbon Nanotube Bundles by Joule Heating', *Applied Physics Letters* 93, no. 2 (2008): 023118 (1-3), https://doi.org/10.1063/1.2957986.
[118] Deuk Seok Chung et al., 'Carbon Nanotube Electron Emitters with a Gated Structure Using Backside Exposure Processes', *Applied Physics Letters* 80, no. 21 (2002): 4045–47, https://doi.org/10.1063/1.1480104.
[119] Young Chul Choi et al., 'The High Contrast Ratio and Fast Response Time of a Liquid Crystal Display Lit by a Carbon Nanotube Field Emission Backlight Unit', *Nanotechnology* 19, no. 23 (2008): 235306 (1-5), https://doi.org/10.1088/0957-4484/19/23/235306.

Structurally CNT is very similar to large rod-like molecule and depicts a liquid crystal (LC)-like behaviour[120]. It was reported that CNTs can be employed to align LC molecules[121] or LC can align CNTs[122]. Moreover, the CNT emits polarized light[123] due to the one-dimensional electron movement through it and the incandescent brightness[124] could reach about 6000 cd/m$^2$. Schindler et al.[125] demonstrated that CNT networks can be used for flexible display as well as FET made with SWCNTs can be used in LCD.

Aguirre et al.[126] synthesised organic light-emitting diodes (OLEDs) using conductive and transparent SWCNT sheets with the luminance efficiency of 1.4 cd A$^{-1}$ and brightness of 2800 cd m$^{-2}$. Li et el.[127] used SWCNT films on flexible substrates as flexible, transparent anodes for OLEDs which depicted a current efficiency of 1.6 cd/A and a maximum light output of 3500 cd/m$^2$. Chein et al.[128] fabricated top emission OLEDs with CNTs as top electrodes with a maximum current efficiency of 1.24 cd A$^{-1}$ and a peak luminance of 3588 cd m$^{-2}$. An efficient OLED was fabricated by Williams et al.[129] employing transparent and strong CNT sheets as the hole-injecting anode which showed current efficiency near 2.5 cd/A and a brightness of 4500 cd/m$^2$. Ou et al.[130] fabricated OLED devices with conductive and transparent CNT anodes with high efficiency of 10 cd/A and luminescence of 9000 cd/m$^2$. Yu et al.[131] used CNTs as cathode and anode to fabricate fully bendable polymer light emitting

---

[120] W. Song, 'Nematic Liquid Crystallinity of Multiwall Carbon Nanotubes', *Science* 302, no. 5649 (2003): 1363, https://doi.org/10.1126/science.1089764.

[121] Joette M. Russell et al., 'Alignment of Nematic Liquid Crystals Using Carbon Nanotube Films', *Thin Solid Films* 509, no. 1–2 (2006): 53–57, https://doi.org/10.1016/j.tsf.2005.09.099.

[122] Jan Lagerwall et al., 'Nanotube Alignment Using Lyotropic Liquid Crystals', *Advanced Materials* 19, no. 3 (2007): 359–64, https://doi.org/10.1002/adma.200600889.

[123] S. B. Singer et al., 'Polarized Light Emission from Individual Incandescent Carbon Nanotubes', *Physical Review B - Condensed Matter and Materials Physics* 83, no. 23 (2011): 233404 (1-4), https://doi.org/10.1103/PhysRevB.83.233404.

[124] Peng Liu et al., 'Fast High-Temperature Response of Carbon Nanotube Film and Its Application as an Incandescent Display', *Advanced Materials* 21, no. 35 (2009): 3563–66, https://doi.org/10.1002/adma.200900473.

[125] Axel Schindler et al., 'Solution-Deposited Carbon Nanotube Layers for Flexible Display Applications', *Physica E: Low-Dimensional Systems and Nanostructures* 37, no. 1–2 (2007): 119–23, https://doi.org/10.1016/j.physe.2006.07.016.

[126] C. M. Aguirre et al., 'Carbon Nanotube Sheets as Electrodes in Organic Light-Emitting Diodes', *Applied Physics Letters* 88, no. 18 (2006): 183104 (1-3), https://doi.org/10.1063/1.2199461.

[127] Jianfeng Li et al., 'Organic Light-Emitting Diodes Having Carbon Nanotube Anodes', *Nano Letters* 6, no. 11 (2006): 2472–77, https://doi.org/10.1021/nl061616a.

[128] Yu Mo Chien et al., 'A Solution Processed Top Emission OLED with Transparent Carbon Nanotube Electrodes', *Nanotechnology* 21, no. 13 (2010): 134020 (1-5), https://doi.org/10.1088/0957-4484/21/13/134020.

[129] Christopher D. Williams et al., 'Multiwalled Carbon Nanotube Sheets as Transparent Electrodes in High Brightness Organic Light-Emitting Diodes', *Applied Physics Letters* 93, no. 18 (2008): 183506 (1-3), https://doi.org/10.1063/1.3006436.

[130] Eric C.W. Ou et al., 'Surface-Modified Nanotube Anodes for High Performance Organic Light-Emitting Diode', *ACS Nano* 3, no. 8 (2009): 2258–64, https://doi.org/10.1021/nn900406n.

[131] Zhibin Yu et al., 'Fully Bendable Polymer Light Emitting Devices with Carbon Nanotubes as Cathode and Anode', *Applied Physics Letters* 95, no. 20 (2009): 203304 (1-3), https://doi.org/10.1063/1.3266869.

devices to achieve high brightness of 1400 cd/m$^2$, low turn-on voltage of 3.8 V and efficiency of 2.2 cd/A. Freitag et al.[132] constructed top-emitting white organic light emitting devices using CNT which showed well balanced nearly Lambertian white emission with colour rendering indices of 70 and negligible spectral change with higher viewing angle. Inigo et al.[133] reported that the presence of COOH-MWCNT at the anode increases hole injection provides higher brightness (20,000 cd/m$^2$) with lower operating voltages. Zhang et al.[134] demonstrated practical realisation of active matrix organic light-emitting diode (AMOLED) displays operated by separated nanotube thin-film transistors (SN-TFTs) incorporating key technological components, viz. bilayer gate dielectric for improved substrate adhesion to the deposited nanotube film, CNT film density optimization and high-yield, large-scale fabrication of devices with better performance. Sekitani et al.[135] fabricated stretchable AMOLED display using printable elastic conductors mainly comprising of SWCNTs. Without creating any mechanical or electrical damage, the display could be spread over a hemisphere and stretched by 30–50%. McCarthy et al.[136] demonstrated an organic channel light-emitting transistor operating with high aperture ratio, low power dissipation and at low voltage, in the three primary colours by employing a SWCNT network electrode that allows integration of the light emitter and the drive transistor into an efficient single stacked device. Different display devices fabricated using CNT are depicted in Fig. 10.

---

[132] P. Freitag et al., 'Lambertian White Top-Emitting Organic Light Emitting Device with Carbon Nanotube Cathode', *Journal of Applied Physics* 112, no. 11 (2012): 114505 (1-5), https://doi.org/10.1063/1.4767439.

[133] A. R. Inigo, J. M. Underwood, and S. R.P. Silva, 'Carbon Nanotube Modified Electrodes for Enhanced Brightness in Organic Light Emitting Devices', *Carbon* 49, no. 13 (2011): 4211–17, https://doi.org/10.1016/j.carbon.2011.05.053.

[134] Jialu Zhang et al., 'Separated Carbon Nanotube Macroelectronics for Active Matrix Organic Light-Emitting Diode Displays', *Nano Letters* 11, no. 11 (2011): 4852–58, https://doi.org/10.1021/nl202695v.

[135] Tsuyoshi Sekitani et al., 'Stretchable Active-Matrix Organic Light-Emitting Diode Display Using Printable Elastic Conductors', *Nature Materials* 8, no. 6 (2009): 494–99, https://doi.org/10.1038/nmat2459.

[136] M. A. McCarthy et al., 'Low-Voltage, Low-Power, Organic Light-Emitting Transistors for Active Matrix Displays', *Science* 332, no. 6029 (2011): 570–73, https://doi.org/10.1126/science.1203052.

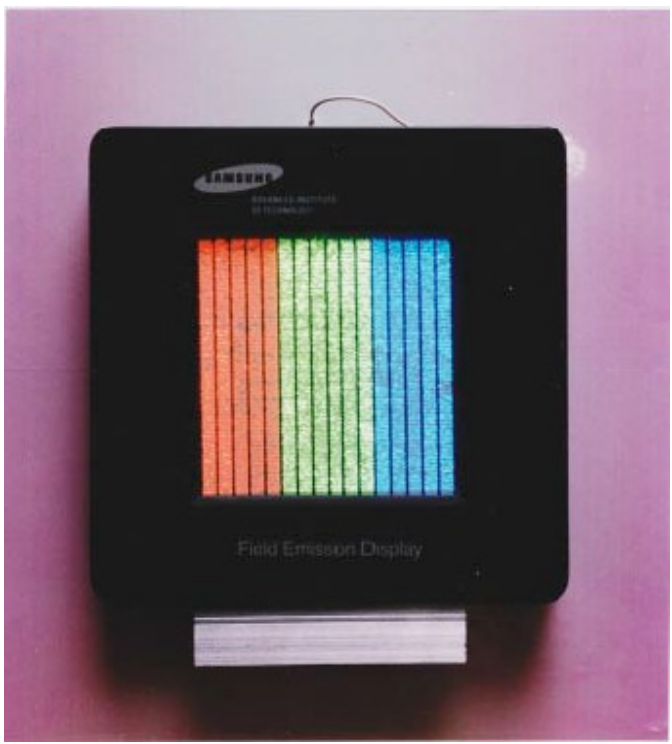

*Fig. 10. Emitting image of fully sealed SWNT-FED at color mode with red, green, and blue phosphor columns[137].*

## 4. Graphene

From the historical perspective, graphene was theoretically predicted as a part of graphite in 1940's[138] and later in 1962 single layer of graphite was first experimentally observed by Boehm et al[139] and afterwards he coined the term **"Graphene"** [140] for the first time in the year of 1994. Nearly two decades later, Novoselov et al.[141] reported the observation of graphene in the journal of Science, which is undoubtedly responsible for the explosion of interest in graphene research in the scientific world and resulted in the fast progress of this field. After this report, graphene has become the most discussed material in the nano material research community. Carbon chains at the edge of graphene are arranged either in zigzag or armchair else in arbitrary fashion as shown in Fig. 11. which is responsible for various conducting behaviours. Metal like behaviour was observed in graphene nanoribbon with a zigzag edge while a nanoribbon with an armchair edge behaves either as metal or semiconductor[142]. Moreover, Graphene nanoribbons (GNR) with homogeneous zigzag or armchair shaped edges have energy gaps which decrease as the width of the nanoribbon is increased[143]. Thus, manipulation of width of the graphene nanoribbon provides precise control over the height of the energy barrier for potential application of graphene in nanoelectronic devices.

---

[137] Choi et al., 'Fully Sealed, High-Brightness Carbon-Nanotube Field-Emission Display'.
[138] P. R. Wallace, 'The Band Theory of Graphite', *Physical Review* 71, no. 9 (1947): 622–34, https://doi.org/10.1103/PhysRev.71.622.
[139] H. P. Boehm et al., 'Dünnste Kohlenstoff-Folien', *Zeitschrift Fur Naturforschung - Section B Journal of Chemical Sciences* 17, no. 3 (1962): 150–53, https://doi.org/10.1515/znb-1962-0302.
[140] H. P. Boehm, R. Setton, and E. Stumpp, 'Nomenclature and Terminology of Graphite Intercalation Compounds (IUPAC Recommendations 1994)', *Pure and Applied Chemistry* 66, no. 9 (1994): 1893–1901, https://doi.org/10.1351/pac199466091893.
[141] Y. Zhang K. S. Novoselov, A. K. Geim, S. V. Morozov, D. Jiang and and A. A. Firsov S. V. Dubonos, I. V. Grigorieva, 'Electric Field Effect in Atomically Thin Carbon Films', *Science* 306, no. 5696 (2004): 666–69, https://doi.org/10.1126/science.1102896.

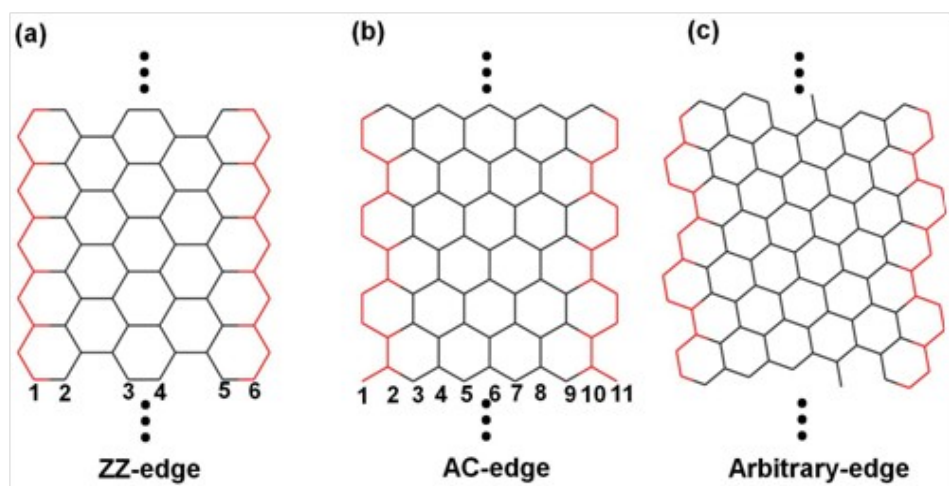

*Fig. 11. Various graphene nanoribbons (GNRs): (a) zigzag GNR (ZGNR), (b) armchair GNR (AGNR) and (c) chrial GNR (CGNR). The number 1, 2, 3,… in (a) and (b) indicate the zigzag and armchair chains of the GNRs and their width[144].(Reprinted with permission from Royal Society of Chemistry)*

**4.1 Graphene application in touch screens**

From the inception of the idea of "touch screen"[145] till date metal oxides[146] have been predominantly used for its fabrication. Among different metal oxides Indium Tin Oxide (ITO)[147] is the most promising due to the exotic combination of visible light transparency

---

and reasonable conductivity[148]. However, owing to the chemical instability of ITO[149], and the high cost[150] resulting from the scarcity of indium and limited flexibility necessitates the search for good alternative. In order to circumvent the challenges imposed on ITO, scientists are in search of a good replacement for ITO and in due course various forms of materials[151] and also nanomaterials[152] have been tested to check the replacement feasibility. The main criterions behind the search are: the new material should be cheaper, chemically stable and more flexible than ITO while the transparency and conductivity of the new material should be better or comparable to ITO. Various studies[153,154,155,156,157,158,159] revealed that graphene is a potential candidate to be used as replacement of ITO because of its low sheet resistance, high optical transmittance, and the possibility of transferring it onto plastic substrates. Furthermore, graphene-based transparent conductive electrodes can be synthesised on a large scale[160] specially by employing the roll-to-roll technique to fabricate touch-screens[161]. Zhu et al.[162] fabricated flexible, transparent, conducting films by employing graphene and a metallic grid (Fig. 12). Transparent electrodes made of the conducting films and transparent substrate

---

[148] O. Mryasov and A. Freeman, 'Electronic Band Structure of Indium Tin Oxide and Criteria for Transparent Conducting Behavior', *Physical Review B* 64, no. 23 (2001): 233111 (1-3), https://doi.org/10.1103/PhysRevB.64.233111.

[149] M. Senthilkumar et al., 'Electrochemical Instability of Indium Tin Oxide (ITO) Glass in Acidic PH Range during Cathodic Polarization', *Materials Chemistry and Physics* 108, no. 2–3 (2008): 403–7, https://doi.org/10.1016/j.matchemphys.2007.10.030.

[150] Peter Harrop, 'Indium Tin Oxide Prices Rocket -Alternative Needed', 2007, https://www.printedelectronicsworld.com/articles/542/indium-tin-oxide-prices-rocket-alternative-needed.

[151] Tadatsugu Minami, 'Substitution of Transparent Conducting Oxide Thin Films for Indium Tin Oxide Transparent Electrode Applications', *Thin Solid Films* 516, no. 7 (2008): 1314–21, https://doi.org/10.1016/j.tsf.2007.03.082.

[152] Michael Layani, Alexander Kamyshny, and Shlomo Magdassi, 'Transparent Conductors Composed of Nanomaterials', *Nanoscale* 6, no. 11 (2014): 5581–91, https://doi.org/10.1039/C4NR00102H.

[153] Akshay Kumar and Chongwu Zhou, 'The Race To Replace Tin-Doped Indium Oxide: Which Material Will Win?', *Acs Nano* 4, no. 1 (2010): 11–14.

[154] Rickard Arvidsson et al., 'Energy and Resource Use Assessment of Graphene as a Substitute for Indium Tin Oxide in Transparent Electrodes', *Journal of Cleaner Production* 132 (2016): 289–97, https://doi.org/10.1016/j.jclepro.2015.04.076.s

[155] Jong Hyun Ahn and Byung Hee Hong, 'Graphene for Displays That Bend', *Nature Nanotechnology* 9, no. 10 (2014): 737–38, https://doi.org/10.1038/nnano.2014.226.

[156] Sang Jin Kim et al., 'Materials for Flexible, Stretchable Electronics: Graphene and 2D Materials', *Annual Review of Materials Research* 45, no. 1 (2015): 63–84, https://doi.org/10.1146/annurev-matsci-070214-020901.

[157] F. Bonaccorso et al., 'Graphene Photonics and Optoelectronics', *Nature Photonics* 4, no. 9 (2010): 611–22, https://doi.org/10.1038/nphoton.2010.186.

[158] Tanmoy Das et al., 'Graphene-Based Flexible and Wearable Electronics', *Journal of Semiconductors* 39, no. 1 (2018): 011007 (1-19), https://doi.org/10.1088/1674-4926/39/1/011007.

[159] Sukang Bae et al., 'Towards Industrial Applications of Graphene Electrodes', *Physica Scripta* T146 (2012): 014024 (1-8), https://doi.org/10.1088/0031-8949/2012/T146/014024.

[160] Keun Soo Kim et al., 'Large-Scale Pattern Growth of Graphene Films for Stretchable Transparent Electrodes', *Nature* 457, no. 7230 (2009): 706–10, https://doi.org/10.1038/nature07719.

[161] Jaechul Ryu et al., 'Fast Synthesis of High-Performance Graphene by Rapid Thermal Chemical Vapor Deposition', *ACS Nano* 8, no. 1 (2014): 950–56, https://doi.org/10.1021/nn405754d.

[162] Yu Zhu, 'Rational Design of Hybrid Graphene Films for High-Performance', *ACS Nano* 5, no. 8 (2011): 6472–79, https://doi.org/10.1021/nn201696g.

(e.g. polyethylene terephthalate or glass) revealed sheet resistance as low as 3 Ω/square with the transmittance at ∼80%. To check the flexibility of the material bending tests were performed which confirmed the hybrid film's conductivity decreases by 20-30% up to 50 bends, while the material then stabilizes and no major variations in conductivity were noticed up to 500 bending cycles. Moreover, the materials employed for the new hybrid electrode are earth-abundant stable elements thereby reducing the cost of the production in comparison to ITO. Jurewicz et al.[163] deposited graphene on ultra-low density networks of AgNW using a Langmuir-based technique thus fabricated transparent, low cost, highly conducting electrodes. Graphene connecting adjacent wires as well as graphene adsorption at inter-wire junctions contributed towards enhanced electrical properties by several orders of magnitude. Moreover, the amount of expensive nanowires required to build such touch screen using pristine high-density nanowire networks is reduced by more than fifty times. The AgNW and graphene composite seems to be a near future replacement of ITO for touch screen application[164].

---

[163] Izabela Jurewicz et al., 'Insulator-Conductor Type Transitions in Graphene-Modified Silver Nanowire Networks: A Route to Inexpensive Transparent Conductors', *Advanced Functional Materials* 24, no. 48 (2014): 7580–87, https://doi.org/10.1002/adfm.201402547.

[164] Matthew J. Large et al., 'Selective Mechanical Transfer Deposition of Langmuir Graphene Films for High-Performance Silver Nanowire Hybrid Electrodes', *Langmuir* 33, no. 43 (2017): 12038–45, https://doi.org/10.1021/acs.langmuir.7b02799.

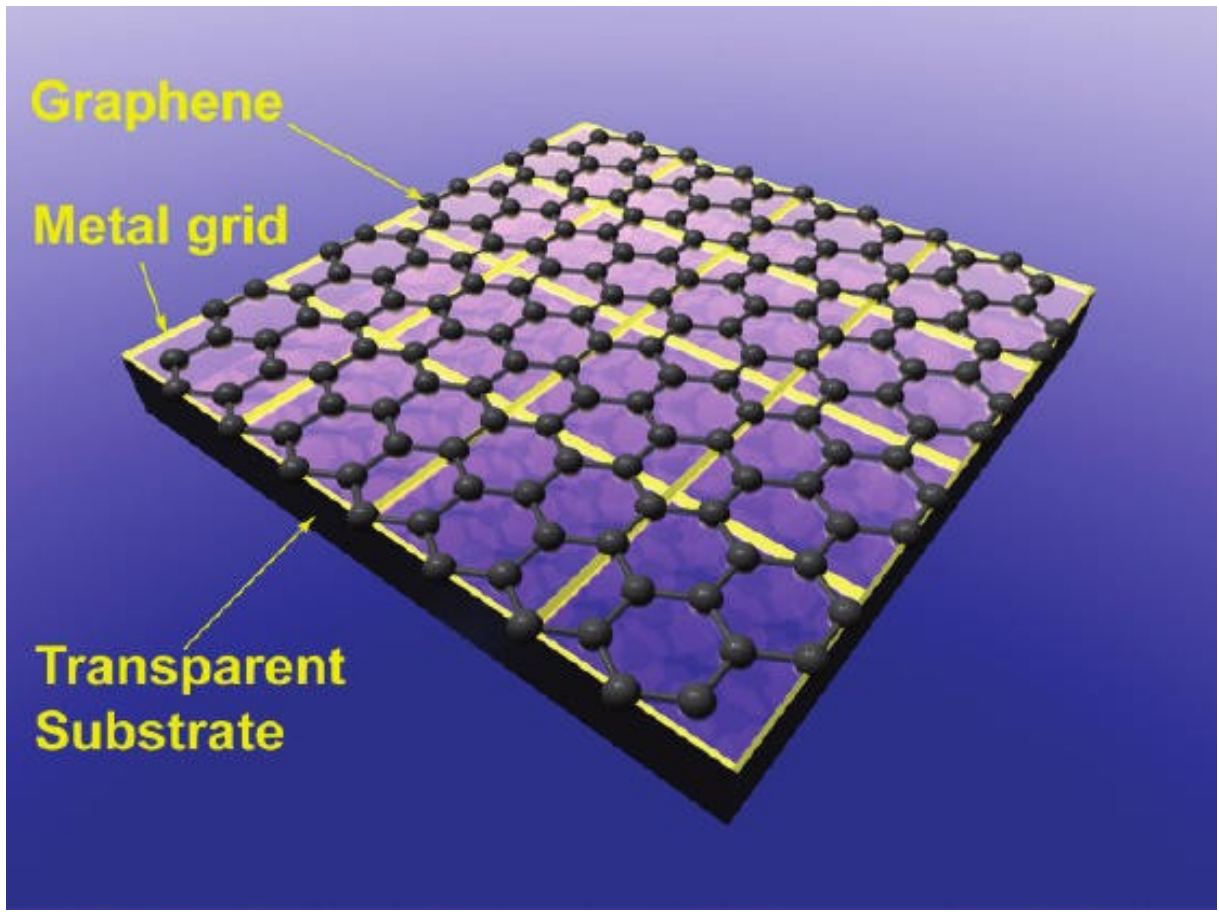

*Fig 12. Metal grid/graphene hybrid transparent electrode. The yellow lines in the figure represent the metal grid. The grid size and gridline width in the figure are only illustrative and are not scaled with the graphene molecular structure. (Reprinted with permission from American Chemical Society)*

**4.2 Graphene application in transparent memory**

Modern electronic gadgets encounter extensive renovations in order to integrate multiple functionalities, viz. wearability, flexibility, portability and transparency. This necessitates the fabrication of core functional segments of devices to be integrated on a single flexible-transparent substrate without reducing the transparency. Initially transparent electronics is fabricated using semiconducting nanowires[165] employing the wide band-gap semiconductors[166]. After the incorporation of graphene into the electronic world, it was observed that graphene can enhance the performance transparent-flexible electronic modules because due to its higher mobility, minimal light absorbance and superior mechanical properties (Fig. 13). Kim et al.[167,168] prepared a flexible and transparent graphene charge-trap memory (GCTM) using a single-layer graphene channel on a polyethylene naphtalate (PEN) substrate below eutectic temperatures (~110 °C). The GCTM loses only 8% transparency compared to PEN substrate and exhibited 8.6 V memory window. Under both compressive and tensile stress, the GCTM reveals minimal perturbation on electrical characteristics. Yao et al.[169] fabricated non-volatile, highly transparent memory devices employing graphene and $SiO_x$, having low programming currents and long retention time. Graphene provides the advantage of flexibility, transparency and potential low cost. With well established $SiO_x$ technology in the semiconductor industry and the precise control over synthesis of graphene promotes the $SiO_x$–graphene memory system as advantages in case of transparent materials composition and processing. Yang et al.[170] used graphene as a stable and transparent resistive element to fabricate a ZnO-based transparent resistive RAM with switching characteristics for memory applications. The significant increase in the insensitivity to the environmental atmosphere and also in the switching yield was observed by statistical analysis, thus graphene plays an important role to suppress the surface effect. Kim et al.[171] used ITO/RGO/ITO structures as transparent electronic memory cell with multi-level resistive switching. The transmittance of memory device was above 80% (including the substrate) in the visible region and by using 2 V to 7 V pulses multi-level resistive switching behaviour was observed. The device showed long data retention of over $10^5$s at 85°C and a good endurance of over $10^5$cycles. Wu et al.[172] fabricated ITO/GO/ITO/PES transparent and flexible resistive switching memory devices with set and reset voltages are around 1 V and 0.1 V, respectively. The resistive switching characteristics of the flexible devices were nearly unaltered after the bending test and the retention time was $10^5$ s. Thus it is quite established that Graphene oxide (GO) resistive memories are inexpensive, eco-friendly with high optical transparency and large mechanical flexibility. Currently, the miniaturisation and enhancement of temporal scalability of GO memories are of keen interest to the scientific community. Recently, Nagareddy et al.[173] fabricated hybrid GO-titanium oxide non-volatile resistive memory with nanometric size-scaling (50 nm), nanosecond switching speeds (sub-5 ns) and multilevel (4-level, 2-bit per cell) operation along with outstanding retention and endurance performance- all on both flexible and rigid substrates. The resistive switching mechanism in the Pt/GO/Ti/Pt devices was controlled via redox reactions in the interfacial region between the GO layer and the top (Ti) electrode.

---
[165] Sanghyun Ju et al., 'Fabrication of Fully Transparent Nanowire Transistors for Transparent and Flexible Electronics', *Nature Nanotechnology* 2, no. 6 (2007): 378–84, https://doi.org/10.1038/nnano.2007.151.
[166] J. F. Wager, D. A. Keszler and R. E. Presley., "Transparent Electronics," Springer Science: New York, (2010)

---

[167] Sung Min Kim et al., 'Transparent and Flexible Graphene Charge-Trap Memory', *ACS Nano* 6, no. 9 (2012): 7879–84, https://doi.org/10.1021/nn302193q.
[168] Sung Min Kim et al., 'Flexible and Transparent Memory: Non-Volatile Memory Based on Graphene Channel Transistor for Flexible and Transparent Electronics Applications', in *2012 4th IEEE International Memory Workshop, IMW 2012*, 2012, 3–6, https://doi.org/10.1109/IMW.2012.6213669.
[169] Jun Yao et al., 'Highly Transparent Nonvolatile Resistive Memory Devices from Silicon Oxide and Graphene', *Nature Communications* 3 (2012): 1101 (1-8), https://doi.org/10.1038/ncomms2110.
[170] Po Kang Yang et al., 'Fully Transparent Resistive Memory Employing Graphene Electrodes for Eliminating Undesired Surface Effects', *Proceedings of the IEEE* 101, no. 7 (2013): 1732–39, https://doi.org/10.1109/JPROC.2013.2260112.
[171] Hee Dong Kim et al., 'Transparent Multi-Level Resistive Switching Phenomena Observed in ITO/RGO/ITO Memory Cells by the Sol-Gel Dip-Coating Method', *Scientific Reports* 4 (2014): 4614 (1-6), https://doi.org/10.1038/srep04614.
[172] Hsiao Yu Wu, Chun Chieh Lin, and Chu Hsuan Lin, 'Characteristics of Graphene-Oxide-Based Flexible and Transparent Resistive Switching Memory', *Ceramics International* 41, no. S1 (2015): S823–28, https://doi.org/10.1016/j.ceramint.2015.03.129.

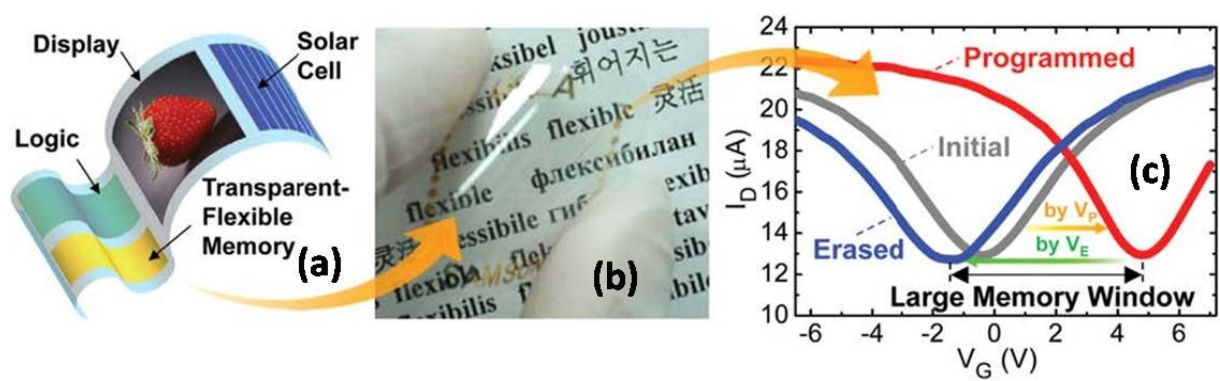

*Fig 13. (a) Flexible transparent memory can be utilized for transparent and flexible electronics that require integration of logic, memory and display on a single substrate with high transparency and endurance under, (b) Photograph of the large scale transparent-flexible flexible transparent memory array fabricated on PEN substrate. The flexible transparent memory cell is transparent enough to see objects through it without image distortion. flex. (c) Memory characteristics of flexible transparent memory. (Reprinted with permission from American Chemical Society)*

## 5. Summary and outlook

---

[173] V. Karthik Nagareddy et al., 'Multilevel Ultrafast Flexible Nanoscale Nonvolatile Hybrid Graphene Oxide-Titanium Oxide Memories', *ACS Nano* 11, no. 3 (2017): 3010–21, https://doi.org/10.1021/acsnano.6b08668.

Significance of carbon Nanoelectronics is stimulated by the exotic properties of carbon nanomaterials containing carbon atoms with sp$^2$ bonding in a hexagonal fashion. The sp$^2$ orbital hybridization introduces an out of plane p$_z$ ($\pi$) orbital which is largely accountable for low-energy transport of electron and three in-plane electronic orbitals mostly accountable for strong C-C bonding. The novelty of these species is evident in the band structure which depicts a linear E-***k*** low-energy band structure with diminishing effective mass and nearly symmetric but extremely high, electron and hole mobility. The robustly bonded hexagonal carbon atoms uphold mechanical toughness even at high temperature and large current condition along with superior thermal conductivity. The thickness of single atomic-layer carbon nanomaterials reduce their susceptibility to short channel effects[174] and are very useful for harsh environments[175]. Finally, carbon nanomaterials are very well suited for electronic applications and deserve major research interests to convert their promise to realization.

---

[174] Aaron D Franklin et al., 'Carbon Nanotube Complementary Wrap-Gate Transistors', *Nano Letters* 13 (2013): 2490−2495, https://doi.org/10.1021/nl400544q.
[175] Cory D. Cress et Al., 'Radiation Effects in Carbon Nanoelectronics', *Electronics* 1 (2012): 23–31, https://doi.org/10.3390/electronics1010023.